\documentclass[aps,prb,twocolumn,floatfix]{revtex4-1}

\pdfoutput=1

\usepackage{times}
\usepackage{amsfonts}
\usepackage{amssymb}
\usepackage{amsmath}
\usepackage{graphicx}
\usepackage{bm}
\usepackage{verbatim}

\usepackage{bm}
\usepackage{color}
\usepackage{stackrel}
\usepackage{accents}
\usepackage{hyperref}
\usepackage{latexsym}
\usepackage{float}
\usepackage{physics}

\usepackage{ulem}

\newcommand{\rr}{\mathbf{r}}
\newcommand{\nn}{\nonumber}

\newcommand{\kk}{\mathbf{k}}
\newcommand{\kq}{\mathbf{q}}
\newcommand{\wn}{{i\omega_n}}
\newcommand{\vm}{{i\nu_m}}
\newcommand{\wh}{\omega_{\text{H}}}
\newcommand{\wbs}{\omega_{\text{BS}}}

\newcommand{\gd}{\gamma_{d,\kk}}



\begin{document}

\title{
Signatures of Bardasis-Schrieffer mode excitation in Third-Harmonic generated currents}

\author{Marvin A. M\"{u}ller and Ilya M. Eremin}
\affiliation{Institut f\"{u}r Theoretische Physik III, Ruhr-Universit\"{a}t Bochum, D-44801 Bochum, Germany }

\begin{abstract}
We theoretically analyze the collective modes in unconventional superconductors focusing on Bardasis-Schrieffer (BS) mode and its contribution to the third harmonic generation currents. Starting from a model with competing superconducting pairing instabilities we add fluctuations of the fields beyond saddle point approximation and calculate their response to an applied pulsed electric field. To model phase fluctuations appropriately we take into account the effect of the long-range Coulomb interaction. While the phase mode is pushed into a plasmon frequency, as known from the literature, we show that the BS mode remains unaffected. Furthermore, it has a characteristic polarization dependence and, unlike the Higgs mode, generates a current in perpendicular direction to the applied field. We find that the Bardasis-Schrieffer excitations contribute a sizable signal to the third harmonic generated current, which is clearly distinguishable from the charge density fluctuations due to Cooper pair breaking effects and can be straightforwardly detected in experiment.
\end{abstract}

\maketitle

\section{Introduction}
The recent technological development of THz spectroscopy makes it possible to probe properties of quantum matter, which cannot be observed in equilibrium. This is of considerable interest in the field of unconventional superconductivity, where controlled probing of the relaxation dynamics yields access to understanding ground state properties of the underlying system.\cite{Averitt2002,Gianetti2016,Shimano_review2020} The THz waves can excite the superconducting state at energies below the quasiparticle continuum. It was found early that in this regime light couples non-linearly to the Cooper pairs and that it excites the collective Higgs mode at $\wh = 2\Delta$\cite{Volkov1974,Amin2004,Barankov2004,Yuzbashyan2005,Yuzbashyan2006,Barankov2006,Papenkort2007,Krull2014,Levchenko2015,Big-Quench-Review2015,Aoki-Higgs2,Matt-Higgs,Cui2019,Schwarz2020,Mootz2020}. This mode corresponds to amplitude oscillation of the superconducting order parameter in the Mexican hat shaped free energy and is therefore also called the amplitude mode. It does not couple to the electromagnetic wave within linear response but becomes visible in the third harmonic generation (THG)\cite{ Matsunaga2014,Tsuji2015,Cea2016,Matsunaga2017,Cea2018,Katsumi2018,Chu2020,Shimano_review2020}.  In particular, below $T_c$ the incident light at some fixed frequency $\Omega$ excites the Higgs mode in a nonlinear process and effectively drives it with  $2\Omega$ during the pulse irradiation. The transmitted light then generates a component which oscillates with the third harmonic of the incident pulse frequency $3\Omega$ due to coupling to this excitation energy. Tuning the effective excitation energy $2\Omega$ to the energy $2\Delta$ then leads to a resonant enhancement of the third harmonic generation. One has to mention, however, that even though it was initially\cite{ Matsunaga2014,Tsuji2015} assumed that the enhancement stems from resonant driving of the Higgs mode frequency $\wh=2\Delta$ it was later shown that the resonance in the clean case is dominated by excitation of charge density fluctuations \cite{Cea2016}, which is also around $2\Delta$, and the contribution to the resonance due to the Higgs mode activation appears to be orders of magnitude smaller. More recently, it was shown that the situation may change in the dirty limit where The Higgs mode can indeed dominate the THG response  \cite{Jujo18,Murotani2019,Silaev2019,Seibold2021,Haenel2021} 

We note by passing that the transition into the superconducting state in conventional superconductors leads to the formation of other modes, including plasmons, and the Carlson-Goldman mode \cite{Basov2005,Sun2020}. The phase (Anderson-Bogoliubov-Goldstone) mode is the order-parameter phase mode, which  couples to the electromagnetic field and in the presence of long-range Coulomb interaction  converts into the plasmon mode\cite{Anderson1963}. In the presence of residual normal state quasiparticles close to T$_c$, the Coulomb potential of the superfluid density fluctuation can be screened, and one finds an ungapped Carlson-Goldman (CG) mode, in which the normal and
superfluid densities oscillate out of phase.\cite{Carlson1975}  

While those types of modes, discussed above, are present in both, conventional and unconventional superconductors, there is another type of collective mode possible in unconventional superconductors. In these systems multiple different pairing symmetries can compete for the superconducting ground state symmetry and if a second pairing symmetry is very close to the ground state symmetry the so-called Bardasis-Schrieffer mode\cite{bardasis61} $\wbs<\wh$ emerges, signaling the nearby subdominant state. 
Its possible experimental observation in the iron-based superconductors due to the close competition between the $s_\pm$ ground state and the nearby $d_{x^2-y^2}$ instability \cite{Kretschmar2013,Boehm2014,wu17,Boehm18,Jost2018,He2020} has triggered further theoretical interest in the properties of this mode\cite{maiti15,maiti16,Allocca2019,mueller18,Mueller19,Sun2020}. Furthermore, it was shown that the nearby nematic instability, if present, couples to the Bardasis-Schrieffer mode and pushes the resulting hybridized Bardasis-Schrieffer-nematic mode further below the quasiparticle continuum and extends the potential observability of this mode beyond the near-degeneracy region of the $d$-wave and $s$-wave superconducting states\cite{Mueller21}
\\
In this manuscript we investigate theoretically the signatures of the Bardasis-Schrieffer mode in the third-harmonic generated current once the driving frequency matches the resonance condition $2\Omega = \wbs$. We show that the strength of the Bardasis-Schrieffer mode signal is of similar strength as the charge density fluctuations even in the clean limit making its observation a straightforward experimental task. The BS mode should yield a strong second resonance alongside the Higgs mode (or charge density fluctuation) frequency. We further study its polarization dependence and show that, unlike the Higgs mode, BS mode generates a current in perpendicular direction to the applied field. 

\section{Effective Action}
Since the current response of a system is given by the variation of the action $S$ with respect to the applied vector potential $\mathbf{j} = -\partial S / \delta \mathbf{A}$ one can obtain the third-harmonic generated current from an action, which is quartic in the vector potential\cite{Cea2016}. Contributions from fluctuating fields couple to the vector potential and thus renormalize the current kernel. 
Our starting point is therefore an action $S = S_0 + S_{sc} + S_{c}$, containing the fermions on a two-dimensional (2D) square lattice $S_0$ interacting via an attractive superconducting interaction $S_{sc}$ and the Coulomb interaction $S_{c}$. The superconducting interaction consists of attractive $s$-wave and $d$-wave channels, respectively, and reads
\begin{align}\label{eq:H_0}
    S_{\text{sc}}= -\int d\tau \sum_{\kk,\kk^\prime,\kq} \left(V_s + V_d\gamma_{\kk,d}\gamma_{\kk^\prime,d}\right)B^\dagger_{\kk,\kq}(\tau) B_{\kk^\prime,\kq}(\tau)  ,
\end{align}
where $V_s$ and $V_d$ are the superconducting interaction strength in the corresponding channels. We choose the $d$-wave form factor as $\gd = \sqrt{2}\cos(2\phi)$, while the $s$-wave interaction is chosen isotropic. Here, we introduce the singlet pair operator $B_{\kk,\kq}(\tau)=  c_{-\kk+\kq/2,\downarrow}(\tau)c_{\kk+\kq/2,\uparrow}(\tau)$ to keep notation simple. Performing a Hubbard-Stratonovic transformation in $ B_{\kk,\kq}$ introduces the superconducting field, which has the form $\Delta_\kk(q) = \Delta_s(q) + \Delta_d(q)\gd$, where $\Delta_s$ is the $s$-wave component transforming like $A_{1g}$ and $\Delta_d$ is the $d$-wave component transforming as $B_{1g}$ with the corresponding form factor.  Note that both $\Delta_s(q)$ and $\Delta_d(q)$ are complex with an arbitrary overall phase. We focus on the $s$-wave ground state by focusing on $V_s/V_d<1$. By performing a gauge transformation $c_{\rr\sigma}\rightarrow c_{\rr\sigma} e^{i \theta(\rr)/2}$ one can choose the ground state field $\Delta_s(q)$ to be real. \\
As mentioned in the Intrduction, the superfluid phase is known to show a sound-like phase (Goldstone) mode $\omega_{\text{G}}  \sim |\kq|$, which couples to the Coulomb field of the lattice and becomes a plasmon\cite{Anderson1963}. This implies that the effect of the Coulomb field needs to be taken into account
\begin{align}
	S_{c} = \int d\tau\sum_{\substack{\kk,\kk^\prime,\kq\\\sigma,\sigma^\prime}}\frac{V_\kq}{2} c^\dagger_{\kk+\kq,\sigma}(\tau)c^\dagger_{\kk^\prime-\kq,\sigma^\prime}(\tau)c_{\kk^\prime,\sigma^\prime}(\tau)c_{\kk,\sigma}(\tau).
\end{align}
Here $V_\kq = 2\pi e^2 / |\kq|$ is the Coulomb potential for charged fermions confined to our 2D lattice. This interaction can be decoupled in the density channel via a Hubbard-Stratonovic transformation introducing the density fluctuations field $\rho(q)$. Finally, the effect of a vector potential can be added to the action via a Peierls substitution $c^\dagger_{\rr,\sigma}c_{\rr+\boldsymbol{\delta},\sigma}\rightarrow e^{ie\mathbf{A}\cdot\boldsymbol{\delta}/c}c^\dagger_{\rr,\sigma}c_{\rr+\boldsymbol{\delta},\sigma}$. 
After a straightforward derivation the total  action acquires the form
\begin{align}
	S = \sum_{k,k^\prime}\Psi^\dagger_{k}\Big[-G_{0}^{-1}(k)\delta_{k,k^\prime}+ \Sigma(k,k^\prime)\Big]\Psi_{k},
\end{align}
where $G_0(k) = \left(\wn\sigma_0 - \xi_\kk\sigma_3  -\Delta_\kk\sigma_1 \right)^{-1}$ is the saddle point Green's function and the self-energy correction $\Sigma(k,k^\prime) = \Sigma_{\Delta_s}(k,k^\prime) + \Sigma_\theta(k,k^\prime) + \Sigma_{\Delta_d^\prime}(k,k^\prime) + \Sigma_{\Delta_d^{\prime\prime}}(k,k^\prime) + \Sigma_\rho(k,k^\prime) + \Sigma_{A_i^2}(k,k^\prime)$, which contains the fluctuations of the corresponding fields $\Delta_s(q),\theta(q),\Delta_d^\prime(q),\Delta^{\prime\prime}_d(q),\rho(q)$ and the vector potential $A^2_i(q)$ around their saddle point value. Here, the $d$-wave superconducting field is separated into real and imaginary parts, $\Delta_d(q) = \Delta_d^\prime(q) -i\Delta_d^{\prime\prime}(q)$. The explicit calculation is shown in Appendix \ref{appendix:a}. We integrate out the fermions to obtain the effective action and keep fluctuations up to quadratic level (Gaussian fluctuations)
\begin{align}
    S_{\text{eff}} =&\nn -\text{Tr}\log\left(G_0^{-1}\right) + \frac{1}{2}\sum_q \boldsymbol{\eta}^T(-q)\hat{\chi}(q)\boldsymbol{\eta}(q)\\ \nn
    &+ \sum_\alpha\eta_{\alpha}^T(-\vm)\chi_{\eta_\alpha,A^2_i}(\vm)A^2_i(\vm)\\
    &+\frac{1}{2}\sum_{i,j}A_i^2(-\vm)K_{0,ij}(\vm)A_j^2(\vm).
\end{align} 
Here, we use the short hand notation for the vector $\boldsymbol{\eta}(q) = \left(\Delta_s(q),\theta(q),\Delta_d^\prime(q),\Delta_d^{\prime\prime}(q),\rho(q)\right)^T$, which includes all fluctuating fields. The corresponding matrix response function, $\hat{\chi}(q)$, is  given by $\chi_{\alpha\beta} = \text{Tr}(G_0\Sigma_{\eta_\alpha}G_0\Sigma_{\eta_\beta})$. In addition, the coupling of the fluctuating fields to the vector potential is mediated via the response functions $\chi_{\eta_\alpha,A^2_i}(\vm)$. The fluctuations of the vector potential itself is mediated via the $2\times 2$ kernel $K_{0,ij} = \text{Tr}(G_0\Sigma_{A^2_i}G_0\Sigma_{A^2_j})$ with $i,j \in \left\{x,y\right\}$. Here, we take the limit $\kq \rightarrow 0$. 

\begin{figure}
	\centering
	\includegraphics[width=\linewidth]{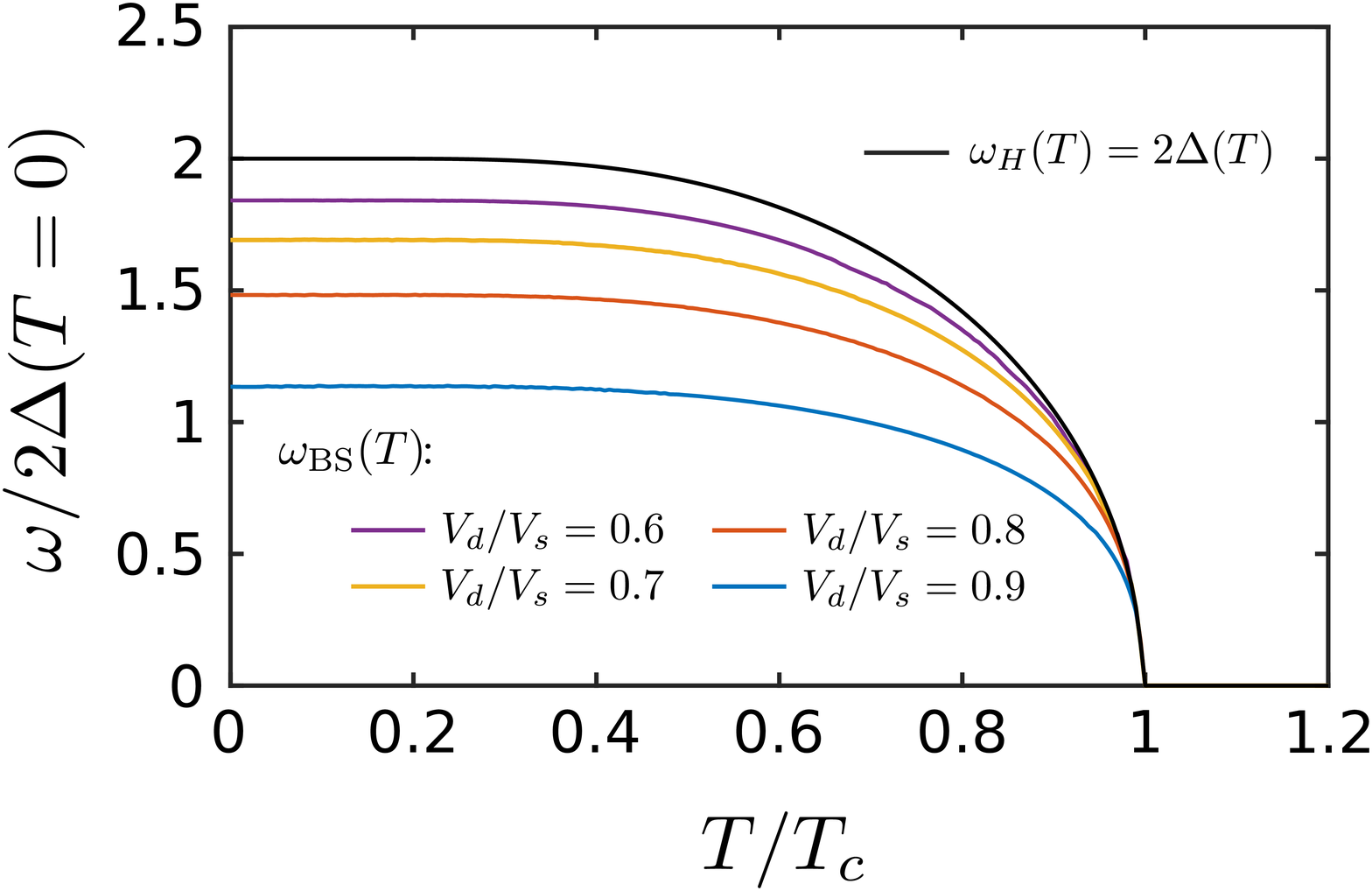}
	\caption{Higgs mode frequency $\wh$ and Bardasis-Schrieffer mode frequency $\wbs$ positions versus temperature $T$ calculated from the resonance frequencies of Eqs. \ref{eq:Higgs} and \ref{eq:BS}. 
	\label{fig:wh_wbs_vs_T}}
\end{figure}

The collective modes of this system are given by the condition $\det(\hat{\chi}(q)) = 0$. In the simplest approximation we neglect the off-diagonal coupling terms and focus on the diagonal terms of $\hat\chi$.
After analytic continuation, one finds for the propagator of the $s$-wave order parameter amplitude
\begin{align}\label{eq:Higgs}
	\chi_{\Delta_{s}\Delta_{s}}(\omega) = \sum_\kk \left(4\Delta-\omega^2\right) F_\kk(\omega)
\end{align}
and the function $F_\kk(\omega) = \tanh(\beta E_\kk/2)/(E_\kk\left(4E^2_\kk-(\omega + i0^+)^2\right))$ carries the information of the Higgs (amplitude) mode $\wh = 2\Delta$. The propagator of the global phase fluctuations is given by
\begin{align}\label{eq:chi_thetatheta_tot}
	\chi_{\theta\theta}(q) =   \frac{1}{4}n_{s}\kq^2 -\omega^2\sum_\kk \Delta^2(\kk)F_\kk(\omega),
\end{align}
where $n_s$ is the superfluid stiffness. This propagator contains the Goldstone (phase) mode, which is gapless and can be excited with an arbitrary small amount of energy. \\
Finally, there is a contribution of the the Bardasis-Schrieffer mode. As was shown previously in Ref. \onlinecite{Mueller21} this mode corresponds to the fluctuations in $\Delta_d^{\prime\prime}$, which, in linear approximation, is the relative phase between the $s$-wave field and the $d$-wave field. The corresponding propagator has the form
\begin{align}\label{eq:BS}
	\chi_{\Delta^{\prime\prime}_d\Delta^{\prime\prime}_d}(\omega) 
	&=\frac{2}{V_d} -\sum_\kk \left(4E^2_\kk \gd^2 \right) F_\kk(\omega).
\end{align}
This function has a single root for $0<\omega<2\Delta$ depending on the exact strength of the $d$-wave interaction $V_d$ relative to $V_s$. Note that the propagator of the amplitude $\Delta_d^\prime$ carries no collective mode at all. In principle the frequency positions of these three modes is slightly affected by the coupling between the fluctuations. However, our analysis shows that the   the cross-coupling between $s$-wave amplitude fluctuations and the global phase $\chi_{\Delta_s\theta}(\omega) =2\omega\sum_\kk \xi_\kk\Delta F_\kk(\omega) $ is present but is very weak. This is similar for the cross-coupling between $\Delta_d^\prime$ and $\Delta_d^{\prime\prime}$. The coupling between the $d$-wave fields $\Delta_d^\prime$ and  $\Delta_d^{\prime\prime}$ and the $s$-wave fields $\Delta_s$ and $\theta$ vanishes because these two channels are orthogonal by symmetry, which implies that these three modes are indeed given by the solution $\chi_{\eta_\alpha\eta_\alpha} = 0$. Correspondingly, in Fig. \ref{fig:wh_wbs_vs_T} the Bardasis-Schrieffer mode and the Higgs mode frequency are shown for different ratios $V_d/V_s$ as a function of temperature $T/T_c$. 
\\
As shown above the Bardasis-Schrieffer mode can be interpreted as a relative phase mode between an $s$-wave field and the $d$-wave field. Therefore a correct treatment of the phase due to incorporating the Coulomb field is  important to describe the Bardasis-Schrieffer mode. Integrating out the charged fields $\rho$ yields the renormalization of the response functions according to
$\chi^r_{AB} = \chi_{AB} - \chi_{\rho A}(-\omega)\chi_{\rho B}(\omega)/\chi_{\rho\rho}(\omega).$
While the effect of this renormalization is weak for the amplitude mode propagator $\chi_{\Delta_s\Delta_s} \simeq \chi^r_{\Delta_s\Delta_s}$, it pushes the Goldstone phase mode into a plasmon %
\begin{align}
\chi^{r}_{\theta\theta}(\kq,\omega) &\simeq \frac{|\kq|}{8\pi e^2}\left(2\pi e^2n_{s}|\kq| - \omega^2\right),
\end{align}
where one can identify the plasmon mode position, $\omega_{Pl} = \sqrt{2\pi e^2n_{s}\kq}$.  The charged field is a density type fluctuations and therefore its fluctuations have the same $A_{1g}$ symmetry as the $s$-wave ground state. Thus, these fluctuations are orthogonal to the subdominant field fluctuations $\Delta_d$ and leave its propagators  and with them the Bardasis-Schrieffer mode, unaffected $\chi^r_{\Delta_d^{\prime/\prime\prime}\Delta_d^{\prime/\prime\prime}} = \chi_{\Delta_d^{\prime/\prime\prime}\Delta_d^{\prime/\prime\prime}}$. Note that while the subdominant field fluctuations fully decouple from the ground state it can be expected that they still give a finite contribution to the third harmonic generated current, as the applied vector potential temporarily breaks $C_4$ rotational symmetry and therefore it allows for a finite mixing in these channels.\\
As mentioned in previous works\cite{Cea2016,Schwarz2020}, the coupling between Higgs mode and the vector potential $\chi_{\Delta_s A_i^2} = \sum_\kk 4\Delta\xi_\kk\frac{\partial^2\xi_\kk}{\partial k_i ^2}F_\kk(\omega)$ is very small as the sum is linear $\xi_\kk$. However, this is not the case for the coupling between Bardasis-Schrieffer mode and the vector potential $\chi_{\Delta_d^{\prime\prime}A^2_i} = \sum_\kk 2i\omega\Delta\gd\frac{\partial^2\xi_\kk}{\partial k_i ^2}F_\kk(\omega)$ and therefore one can expect that unlike the Higgs mode, this mode is easily observable even in the clean limit, where the charge density fluctuations, which are given by $K_{ij} =  -\sum_\kk   4\Delta^2 \frac{\partial^2\xi_\kk}{\partial k_i ^2}\frac{\partial^2\xi_\kk}{\partial k_j ^2}F_\kk(\omega)$ dominates the intensity of the third harmonic generation current.

\section{Third-harmonic response}

The current kernel $\hat{K} = \hat{K}_0 + \hat{K}_{\rho} + \hat{K}_{\Delta_s} + \hat{K}_\theta + \hat{K}_{\Delta_d^\prime} + \hat{K}_{\Delta_d^{\prime\prime}}$ now contains contributions due to each field and in the following we compute the current $j_i(t) = - \frac{\delta S}{\delta A_i(t)}$. We assume that the vector potential can be modelled by  harmonic driving $\mathbf{A}(t) = \mathbf{A}_0\cos(\Omega t)$ with the driving frequency $\Omega$. Using $\phi$ as the polar angle in the momentum space with respect to the $k_x$ axis we write $\mathbf{A}_0 = A_0 \left(\cos(\phi),\sin(\phi)\right)^T$. Here, $A_0$ is the strength of the driving field and the angle $\theta$ denotes the polarization direction. The third-harmonic generation current can be expressed as
\begin{align}\label{eq:current}
    j_{3,i}(3\Omega) &= \int dt j_{3,i}(t)e^{-3i\Omega t} \nonumber \\
    &=\frac{1}{8}\left(\frac{e^2}{2}\right)^2 A_{i,0} \sum_j K_{ij}(2\Omega)A^2_{j,0}
\end{align}
with components in multiple directions depending on the components of the kernel $K_{ij}$.  Therefore, it is useful to introduce the vectors $\mathbf{n}_{\parallel} = \left(\cos(\theta),\sin(\phi)\right)^T$ and  $\mathbf{n}_{\perp} = \left(-\sin(\phi),\cos(\phi)\right)^T$ to filter out the parallel and perpendicular component of the induced current. Although the kernel has in total four components, only two of them are independent by symmetry, i.e. we write $K_{yy} = K_{xx}$ and $K_{yx} = K_{xy}$. Thus, one finds for the parallel and the perpendicular components of the induced current
\begin{align}
    j_{3,\parallel}(\phi) =&  \mathbf{j}_3\cdot \hat{\mathbf{n}} \nn \\=&\frac{1}{8} \left(\frac{e^2}{2}\right)^2\Bigg[\left(\cos^4(\phi) + \sin^4(\phi)\right)K_{xx}(2\Omega) \nn \\ &+ \frac{1}{2}K_{xy}(2\Omega)\sin^2(2\phi)\Bigg] \label{eq:j_para}
    \\\nn
j_{3,\perp}(\phi) =& \mathbf{j}_3\cdot \hat{\mathbf{n}}_{\perp} \\ =& \frac{1}{8} \left(\frac{e^2}{2}\right)^2\left[\frac{1}{4}\sin(4\phi)\left(K_{xy}(2\Omega) - K_{xx}(2\Omega)\right) \right].\label{eq:j_perp}
\end{align}
Before explicitly evaluating these expressions numerically, we summarize the polarization dependence of each excitation, {\it i.e.} the charge density fluctuations (CDF), the Higgs and BS modes as well as phase fluctuations in Table \ref{table:Table_polarization} and the expressions for $K_{ij}$ are explicitly derived in Appendix \ref{appendix:b}. As was shown in Refs.\cite{Schwarz2020,Cea2018} the contribution due to the Higgs or the phase fluctuations show different polarization dependence than the CDF contribution. In fact the polarization dependence can be easily read off if one knows the ratio $K_{xy}/K_{xx}$. Since the $s$-wave amplitude fluctuations, the global phase fluctuations and the density fluctuations have to be $A_{1g}$ symmetric, one finds that  $K_{xy}=K_{xx}$. From Eqs. \ref{eq:j_para} and \ref{eq:j_perp} this implies that those three modes yield no contribution to the perpendicular current and a constant  in $\phi$ contribution to the parallel current. This is different for the contribution of the $d$-wave fields  $\Delta_d^\prime$ and $\Delta_d^{\prime\prime}$. One finds  in this case $K_{xx} = - K_{yy}$, leading to a very different polarization dependence of their contribution to the current. In particular, we obtain that the parallel current has $\cos^2(2\phi)$ polarization dependence, while the perpendicular current has $\sin(4\phi)$ dependence. Thus, the
Bardasis-Schrieffer mode, mediated via fluctuations of the field $\Delta_d^{\prime\prime}$ yields no signal for a periodic driving field direction along the Brillouin zone diagonal $\phi = \pi/4$. This agrees with a previous theoretical analysis of the pump-probe photoemission \cite{Mueller19}. Note, the amplitude fluctuations in the subdominant $d$-wave channel, {\it i.e.} in  $\Delta_d^\prime$, are generally small.
For the CDF contribution there is no strict relation between $K_{xy}$ and $K_{xx}$ and their exact ratio depends on the precise band structure. Therefore CDF yield a mixed polarization profile to the current along the parallel direction. Similar to the Bardasis-Schrieffer mode the CDF shows $\sin(4\phi)$ dependence for the perpendicular current $j_{3,\perp}$. 
\ref{table:Table_polarization}.
\begin{table}
\begin{tabular}{|c||c|c|}
	\hline 
	& $j_{3,\parallel}(\phi)$ & $j_{3,\perp}(\phi)$ \\ 
	\hline \hline
	Higgs mode  & const. & 0 \\ 
	\hline 
	Phase fluctuations & const. & 0  \\ 
	\hline 
	Bardasis-Schrieffer mode & $~\cos^2(2\phi)$ &  $~\sin(4\phi)$ \\ 
	\hline 
	charge density fluctuations & mixed const. $+$  $~\cos^2(2\phi)$  &  $~\sin(4\phi)$ \\ 
	\hline 
\end{tabular} \caption{\label{table:Table_polarization} Summary of the polarization dependence of each contribution to the induced third-harmonic generation current for the parallel and the perpendicular orientation to the applied vector potential.}
\end{table}

It is important to notice that the polarization profile of each contribution is not affected by the renormalization of the propagators by the Coulomb field, as they follow the intrinsic symmetry properties of the fields. Instead, the effect of the renormalization is visible in the explicit dependence on the driving frequency $\Omega$. Observe also that although we assumed the isotropic order parameter in the $s$-wave ground state, our results hold for the general $A_{1g}$-symmetric ground state (like anisotropic $s$-wave) as they follow from the properties of the $A_{1g}$ and $B_{1g}$ irreducible representations under rotation by $\pi/2$ angle. Thus, the presence of a Bardasis-Schrieffer mode signal can be easily detected by the analysis of the polarization dependence of the current. A signal, which is present at $\phi = 0$ but absent at $\phi=\pi/4$ should be a strong indication of the Bardasis-Schrieffer mode (or Bardadsis-Schrieffer nematic mode) and clearly distinguishable from other types of modes. \\
\begin{figure}
	\centering
	\includegraphics[width=\linewidth]{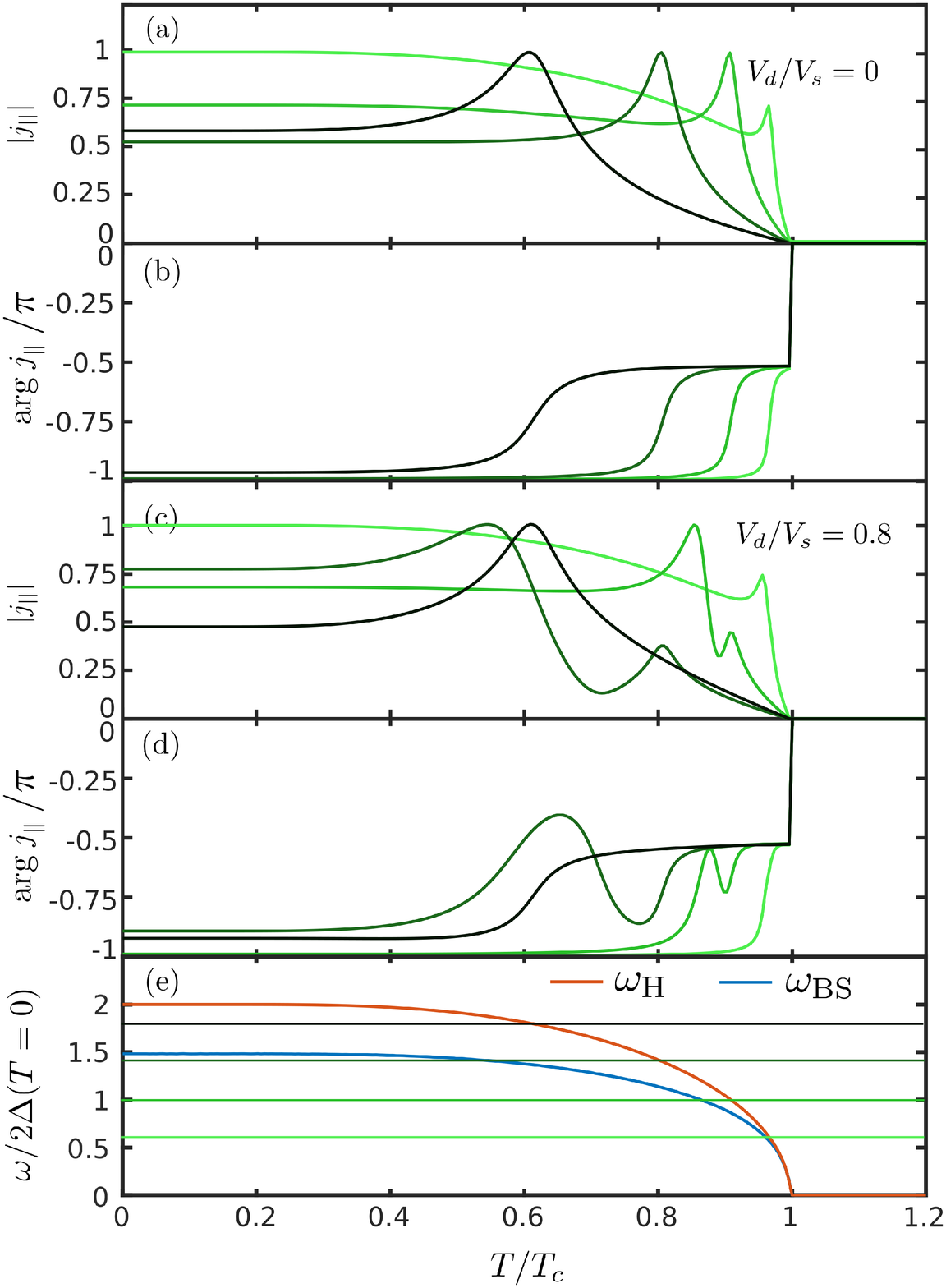}
	\caption{\label{fig:jpara}	The total third-harmonic generated current $j_{3,\parallel}$ parallel to the applied field versus temperature $T$ for $V_d/V_s = 0.8$. The signal is shown without (a) and with (c) taking into account the sub-dominant $d$-wave interaction for the four frequency cuts shown in (e). Additionally the corresponding phase dependence of the current in (b) and (d) are shown.}
\end{figure}
In particular, in Fig. \ref{fig:jpara} we show the third harmonic generated current $j_{3,\parallel}$ parallel to the applied vector potential $\mathbf{A}(t)$ for a system with ($V_d/V_s = 0.8$) and without ($V_d/V_s = 0$) a subdominant $d$-wave instability. Indeed, one finds that the contribution due to the Bardasis-Schrieffer mode is sizable such that apart from the pair breaking signal at $\omega = 2\Delta$ a second resonance condition can be found, which agrees well with the calculated frequencies for the Bardasis-Schrieffer mode. For a constant driving frequency $\Omega$ the resonance due to the Bardasis-Schrieffer mode frequency $\wbs$ is activated at lower temperatures than the resonance at $2\Delta$ and since the current is generally larger for a larger order parameter $\Delta(T)$, this makes the resonance peak at $\omega = \wbs$ stronger than the resonance peak at $\omega = 2\Delta$. Note that in agreement with Ref. \onlinecite{Cea2016} we find that the contribution due to the Higgs mode is small compared to the CDF and the phase contribution. Therefore, the Higgs mode contribution to the total current remains negligible compared to the total current.\\ Similar to the amplitude of the third harmonic generated current, we find strong signatures of the Bardasis-Schrieffer mode in the phase of the current. Due to the two resonance frequencies the phase of the current varies strongly with temperatures in a region between the resonance at $\wbs$ and $2\Delta$. Therefore it appears that the presence of two resonant modes is even more pronounced in the phase of the third harmonic generated current than in the intensity of the signal itself.\\
\begin{figure}
	\centering
	\includegraphics[width=\linewidth]{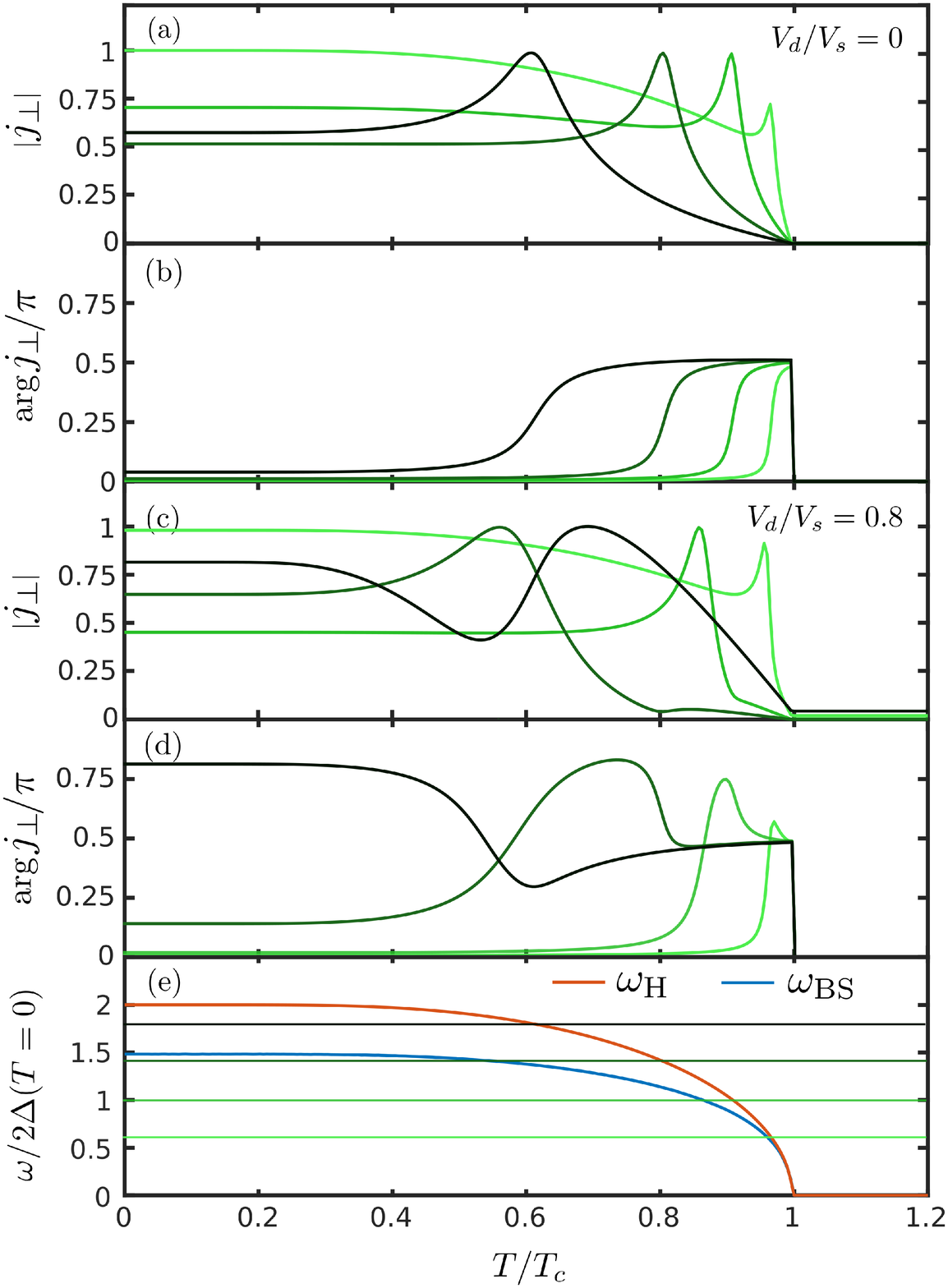}
	\caption{\label{fig:jperp} The total third-harmonic generated current $j_{3,\perp}$ perpendicular to the applied field versus temperature for $V_d/V_s = 0$ (a)-(b) and $V_d/V_s = 0.8$ (c)-(d) for the four frequency cuts shown in (e).}
\end{figure}
The third harmonic generated current induced in perpendicular direction to the vector potential $\mathbf{A}$ is shown in Fig. \ref{fig:jpara}.  In the perpendicular current the signal is dominated by the Bardasis-Schrieffer mode peak once the subdominant channel is present ($V_d/V_s = 0.8$) and is dominated by the pair breaking peak (CDF) at $2\Delta$ if no subdominant channel is present ($V_d/V_s = 0.0$). Similar to the third-harmonic generated current in the parallel direction the presence of a second resonance peak can be clearly visible  in the phase of the current. As only the charge density fluctuations and the fluctuations of the $d$-wave field contribute to the perpendicular current the renormalization effects due to the long-range Coulomb interaction do not influence this component of the current. Although the different contributions to the third harmonic generated current induced parallel to the field can be distinguished by their polarization dependence, this is not the case for the current induced in the perpendicular direction, where all contributions show the same $\sin(4\phi)$ dependence.\\

\section{Conclusion}

To conclude we analyzed theoretically the signatures of the Bardasis-Schrieffer mode excitation in the third harmonic generated currents. Including the long-range Coulomb interaction to ensure a correct treatment of the phase fluctuations, we showed that the Bardasis-Schrieffer mode excitation are clearly visible as a resonance in the third harmonic generated current. Unlike the Higgs mode signal, which is orders of magntiude smaller in the clean limit than the charge density fluctuations  contribution due to the Cooper pair breaking, we demonstrated that the Bardasis-Schrieffer mode signal is of similar strength and can be clearly visible in the magnitude as well as the phase of the current. We further showed that the contribution of the BS mode to the polarization dependence of the third harmonic generated current $\mathbf{j}_3$ has very characteristic features, different from the Higgs, charge-density fluctuations and phase fluctuations modes. This polarization dependence can serve as a smoking gun for the experimental observation of this mode. Furthermore, the Bardasis-Schrieffer mode contributes also to the perpendicular component of the third harmonic generated current and its intensity is also of similar magnitude as the charge density fluctuations. These results clearly open the perspective to observe this mode in unconventional superconductors.

\section{Acknowledgments}
We thank Lara Benfatto, Yan Gallais, Indranil Paul, Anatoly F. Volkov, and Pavel Volkov for useful conversations. The work was supported by the joint NSFC-DFG grant (ER 463/14-1)\\

\bibliography{bibliography}

\begin{thebibliography}{50}%
\makeatletter
\providecommand \@ifxundefined [1]{%
 \@ifx{#1\undefined}
}%
\providecommand \@ifnum [1]{%
 \ifnum #1\expandafter \@firstoftwo
 \else \expandafter \@secondoftwo
 \fi
}%
\providecommand \@ifx [1]{%
 \ifx #1\expandafter \@firstoftwo
 \else \expandafter \@secondoftwo
 \fi
}%
\providecommand \natexlab [1]{#1}%
\providecommand \enquote  [1]{``#1''}%
\providecommand \bibnamefont  [1]{#1}%
\providecommand \bibfnamefont [1]{#1}%
\providecommand \citenamefont [1]{#1}%
\providecommand \href@noop [0]{\@secondoftwo}%
\providecommand \href [0]{\begingroup \@sanitize@url \@href}%
\providecommand \@href[1]{\@@startlink{#1}\@@href}%
\providecommand \@@href[1]{\endgroup#1\@@endlink}%
\providecommand \@sanitize@url [0]{\catcode `\\12\catcode `\$12\catcode
  `\&12\catcode `\#12\catcode `\^12\catcode `\_12\catcode `\%12\relax}%
\providecommand \@@startlink[1]{}%
\providecommand \@@endlink[0]{}%
\providecommand \url  [0]{\begingroup\@sanitize@url \@url }%
\providecommand \@url [1]{\endgroup\@href {#1}{\urlprefix }}%
\providecommand \urlprefix  [0]{URL }%
\providecommand \Eprint [0]{\href }%
\providecommand \doibase [0]{http://dx.doi.org/}%
\providecommand \selectlanguage [0]{\@gobble}%
\providecommand \bibinfo  [0]{\@secondoftwo}%
\providecommand \bibfield  [0]{\@secondoftwo}%
\providecommand \translation [1]{[#1]}%
\providecommand \BibitemOpen [0]{}%
\providecommand \bibitemStop [0]{}%
\providecommand \bibitemNoStop [0]{.\EOS\space}%
\providecommand \EOS [0]{\spacefactor3000\relax}%
\providecommand \BibitemShut  [1]{\csname bibitem#1\endcsname}%
\let\auto@bib@innerbib\@empty
\bibitem [{\citenamefont {Averitt}\ and\ \citenamefont
  {Taylor}(2002)}]{Averitt2002}%
  \BibitemOpen
  \bibfield  {author} {\bibinfo {author} {\bibfnamefont {R.}~\bibnamefont
  {Averitt}}\ and\ \bibinfo {author} {\bibfnamefont {A.~J.}\ \bibnamefont
  {Taylor}},\ }\href@noop {} {\bibfield  {journal} {\bibinfo  {journal}
  {Journal of Physics: Condensed Matter}\ } (\bibinfo {year}
  {2002})}\BibitemShut {NoStop}%
\bibitem [{\citenamefont {Giannetti}\ \emph {et~al.}(2016)\citenamefont
  {Giannetti}, \citenamefont {Capone}, \citenamefont {Fausti}, \citenamefont
  {Fabrizio}, \citenamefont {Parmigiani},\ and\ \citenamefont
  {Mihailovic}}]{Gianetti2016}%
  \BibitemOpen
  \bibfield  {author} {\bibinfo {author} {\bibfnamefont {C.}~\bibnamefont
  {Giannetti}}, \bibinfo {author} {\bibfnamefont {M.}~\bibnamefont {Capone}},
  \bibinfo {author} {\bibfnamefont {D.}~\bibnamefont {Fausti}}, \bibinfo
  {author} {\bibfnamefont {M.}~\bibnamefont {Fabrizio}}, \bibinfo {author}
  {\bibfnamefont {F.}~\bibnamefont {Parmigiani}}, \ and\ \bibinfo {author}
  {\bibfnamefont {D.}~\bibnamefont {Mihailovic}},\ }\href {\doibase
  10.1080/00018732.2016.1194044} {\bibfield  {journal} {\bibinfo  {journal}
  {Advances in Physics}\ }\textbf {\bibinfo {volume} {65}},\ \bibinfo {pages}
  {58} (\bibinfo {year} {2016})},\ \Eprint
  {http://arxiv.org/abs/https://doi.org/10.1080/00018732.2016.1194044}
  {https://doi.org/10.1080/00018732.2016.1194044} \BibitemShut {NoStop}%
\bibitem [{\citenamefont {Shimano}\ and\ \citenamefont
  {Tsuji}(2020)}]{Shimano_review2020}%
  \BibitemOpen
  \bibfield  {author} {\bibinfo {author} {\bibfnamefont {R.}~\bibnamefont
  {Shimano}}\ and\ \bibinfo {author} {\bibfnamefont {N.}~\bibnamefont
  {Tsuji}},\ }\href {\doibase 10.1146/annurev-conmatphys-031119-050813}
  {\bibfield  {journal} {\bibinfo  {journal} {Annual Review of Condensed Matter
  Physics}\ }\textbf {\bibinfo {volume} {11}},\ \bibinfo {pages} {103}
  (\bibinfo {year} {2020})},\ \Eprint
  {http://arxiv.org/abs/https://doi.org/10.1146/annurev-conmatphys-031119-050813}
  {https://doi.org/10.1146/annurev-conmatphys-031119-050813} \BibitemShut
  {NoStop}%
\bibitem [{\citenamefont {Volkov}\ and\ \citenamefont
  {Kogan}(1974)}]{Volkov1974}%
  \BibitemOpen
  \bibfield  {author} {\bibinfo {author} {\bibfnamefont {A.~F.}\ \bibnamefont
  {Volkov}}\ and\ \bibinfo {author} {\bibfnamefont {S.~M.}\ \bibnamefont
  {Kogan}},\ }\href@noop {} {\bibfield  {journal} {\bibinfo  {journal} {Sov.
  Phys. JETP}\ }\textbf {\bibinfo {volume} {38}},\ \bibinfo {pages} {1018}
  (\bibinfo {year} {1974})}\BibitemShut {NoStop}%
\bibitem [{\citenamefont {Amin}\ \emph {et~al.}(2004)\citenamefont {Amin},
  \citenamefont {Bezuglyi}, \citenamefont {Kijko},\ and\ \citenamefont
  {Omelyanchouk}}]{Amin2004}%
  \BibitemOpen
  \bibfield  {author} {\bibinfo {author} {\bibfnamefont {M.}~\bibnamefont
  {Amin}}, \bibinfo {author} {\bibfnamefont {E.}~\bibnamefont {Bezuglyi}},
  \bibinfo {author} {\bibfnamefont {A.}~\bibnamefont {Kijko}}, \ and\ \bibinfo
  {author} {\bibfnamefont {A.}~\bibnamefont {Omelyanchouk}},\ }\href@noop {}
  {\bibfield  {journal} {\bibinfo  {journal} {Low Temp. Phys.}\ }\textbf
  {\bibinfo {volume} {30}},\ \bibinfo {pages} {661} (\bibinfo {year}
  {2004})}\BibitemShut {NoStop}%
\bibitem [{\citenamefont {Barankov}\ \emph {et~al.}(2004)\citenamefont
  {Barankov}, \citenamefont {Levitov},\ and\ \citenamefont
  {Spivak}}]{Barankov2004}%
  \BibitemOpen
  \bibfield  {author} {\bibinfo {author} {\bibfnamefont {R.~A.}\ \bibnamefont
  {Barankov}}, \bibinfo {author} {\bibfnamefont {L.~S.}\ \bibnamefont
  {Levitov}}, \ and\ \bibinfo {author} {\bibfnamefont {B.~Z.}\ \bibnamefont
  {Spivak}},\ }\href {\doibase 10.1103/PhysRevLett.93.160401} {\bibfield
  {journal} {\bibinfo  {journal} {Phys. Rev. Lett.}\ }\textbf {\bibinfo
  {volume} {93}},\ \bibinfo {pages} {160401} (\bibinfo {year}
  {2004})}\BibitemShut {NoStop}%
\bibitem [{\citenamefont {Yuzbashyan}\ \emph {et~al.}(2005)\citenamefont
  {Yuzbashyan}, \citenamefont {Altshuler}, \citenamefont {Kuznetsov},\ and\
  \citenamefont {Enolskii}}]{Yuzbashyan2005}%
  \BibitemOpen
  \bibfield  {author} {\bibinfo {author} {\bibfnamefont {E.}~\bibnamefont
  {Yuzbashyan}}, \bibinfo {author} {\bibfnamefont {B.}~\bibnamefont
  {Altshuler}}, \bibinfo {author} {\bibfnamefont {V.}~\bibnamefont
  {Kuznetsov}}, \ and\ \bibinfo {author} {\bibfnamefont {V.~Z.}\ \bibnamefont
  {Enolskii}},\ }\href@noop {} {\bibfield  {journal} {\bibinfo  {journal}
  {Phys. Rev. B}\ }\textbf {\bibinfo {volume} {72}},\ \bibinfo {pages} {220503}
  (\bibinfo {year} {2005})}\BibitemShut {NoStop}%
\bibitem [{\citenamefont {Yuzbashyan}\ \emph {et~al.}(2006)\citenamefont
  {Yuzbashyan}, \citenamefont {Tsyplyatyev},\ and\ \citenamefont
  {Altshuler}}]{Yuzbashyan2006}%
  \BibitemOpen
  \bibfield  {author} {\bibinfo {author} {\bibfnamefont {E.}~\bibnamefont
  {Yuzbashyan}}, \bibinfo {author} {\bibfnamefont {O.}~\bibnamefont
  {Tsyplyatyev}}, \ and\ \bibinfo {author} {\bibfnamefont {B.~L.}\ \bibnamefont
  {Altshuler}},\ }\href@noop {} {\bibfield  {journal} {\bibinfo  {journal}
  {Phys. Rev. Lett.}\ }\textbf {\bibinfo {volume} {96}},\ \bibinfo {pages}
  {097005} (\bibinfo {year} {2006})}\BibitemShut {NoStop}%
\bibitem [{\citenamefont {Barankov}\ and\ \citenamefont
  {Levitov}(2006)}]{Barankov2006}%
  \BibitemOpen
  \bibfield  {author} {\bibinfo {author} {\bibfnamefont {R.~A.}\ \bibnamefont
  {Barankov}}\ and\ \bibinfo {author} {\bibfnamefont {L.~S.}\ \bibnamefont
  {Levitov}},\ }\href {\doibase 10.1103/PhysRevLett.96.230403} {\bibfield
  {journal} {\bibinfo  {journal} {Phys. Rev. Lett.}\ }\textbf {\bibinfo
  {volume} {96}},\ \bibinfo {pages} {230403} (\bibinfo {year}
  {2006})}\BibitemShut {NoStop}%
\bibitem [{\citenamefont {Papenkort}\ \emph {et~al.}(2007)\citenamefont
  {Papenkort}, \citenamefont {Axt},\ and\ \citenamefont
  {Kuhn}}]{Papenkort2007}%
  \BibitemOpen
  \bibfield  {author} {\bibinfo {author} {\bibfnamefont {T.}~\bibnamefont
  {Papenkort}}, \bibinfo {author} {\bibfnamefont {V.}~\bibnamefont {Axt}}, \
  and\ \bibinfo {author} {\bibfnamefont {T.}~\bibnamefont {Kuhn}},\ }\href@noop
  {} {\bibfield  {journal} {\bibinfo  {journal} {Phys. Rev. B}\ }\textbf
  {\bibinfo {volume} {76}},\ \bibinfo {pages} {224522} (\bibinfo {year}
  {2007})}\BibitemShut {NoStop}%
\bibitem [{\citenamefont {Krull}\ \emph {et~al.}(2014)\citenamefont {Krull},
  \citenamefont {Manske}, \citenamefont {Uhrig},\ and\ \citenamefont
  {Schnyder}}]{Krull2014}%
  \BibitemOpen
  \bibfield  {author} {\bibinfo {author} {\bibfnamefont {H.}~\bibnamefont
  {Krull}}, \bibinfo {author} {\bibfnamefont {D.}~\bibnamefont {Manske}},
  \bibinfo {author} {\bibfnamefont {G.~S.}\ \bibnamefont {Uhrig}}, \ and\
  \bibinfo {author} {\bibfnamefont {A.~P.}\ \bibnamefont {Schnyder}},\ }\href
  {\doibase 10.1103/PhysRevB.90.014515} {\bibfield  {journal} {\bibinfo
  {journal} {Phys. Rev. B}\ }\textbf {\bibinfo {volume} {90}},\ \bibinfo
  {pages} {014515} (\bibinfo {year} {2014})}\BibitemShut {NoStop}%
\bibitem [{\citenamefont {Dzero}\ \emph {et~al.}(2015)\citenamefont {Dzero},
  \citenamefont {Khodas},\ and\ \citenamefont {Levchenko}}]{Levchenko2015}%
  \BibitemOpen
  \bibfield  {author} {\bibinfo {author} {\bibfnamefont {M.}~\bibnamefont
  {Dzero}}, \bibinfo {author} {\bibfnamefont {M.}~\bibnamefont {Khodas}}, \
  and\ \bibinfo {author} {\bibfnamefont {A.}~\bibnamefont {Levchenko}},\ }\href
  {\doibase 10.1103/PhysRevB.91.214505} {\bibfield  {journal} {\bibinfo
  {journal} {Phys. Rev. B}\ }\textbf {\bibinfo {volume} {91}},\ \bibinfo
  {pages} {214505} (\bibinfo {year} {2015})}\BibitemShut {NoStop}%
\bibitem [{\citenamefont {Yuzbashyan}\ \emph {et~al.}(2015)\citenamefont
  {Yuzbashyan}, \citenamefont {Dzero}, \citenamefont {Gurarie},\ and\
  \citenamefont {Foster}}]{Big-Quench-Review2015}%
  \BibitemOpen
  \bibfield  {author} {\bibinfo {author} {\bibfnamefont {E.~A.}\ \bibnamefont
  {Yuzbashyan}}, \bibinfo {author} {\bibfnamefont {M.}~\bibnamefont {Dzero}},
  \bibinfo {author} {\bibfnamefont {V.}~\bibnamefont {Gurarie}}, \ and\
  \bibinfo {author} {\bibfnamefont {M.~S.}\ \bibnamefont {Foster}},\ }\href
  {\doibase 10.1103/PhysRevA.91.033628} {\bibfield  {journal} {\bibinfo
  {journal} {Phys. Rev. A}\ }\textbf {\bibinfo {volume} {91}},\ \bibinfo
  {pages} {033628} (\bibinfo {year} {2015})}\BibitemShut {NoStop}%
\bibitem [{\citenamefont {Murotani}\ \emph {et~al.}(2017)\citenamefont
  {Murotani}, \citenamefont {Tsuji},\ and\ \citenamefont {Aoki}}]{Aoki-Higgs2}%
  \BibitemOpen
  \bibfield  {author} {\bibinfo {author} {\bibfnamefont {Y.}~\bibnamefont
  {Murotani}}, \bibinfo {author} {\bibfnamefont {N.}~\bibnamefont {Tsuji}}, \
  and\ \bibinfo {author} {\bibfnamefont {H.}~\bibnamefont {Aoki}},\ }\href
  {\doibase 10.1103/PhysRevB.95.104503} {\bibfield  {journal} {\bibinfo
  {journal} {Phys. Rev. B}\ }\textbf {\bibinfo {volume} {95}},\ \bibinfo
  {pages} {104503} (\bibinfo {year} {2017})}\BibitemShut {NoStop}%
\bibitem [{\citenamefont {Chou}\ \emph {et~al.}(2017)\citenamefont {Chou},
  \citenamefont {Liao},\ and\ \citenamefont {Foster}}]{Matt-Higgs}%
  \BibitemOpen
  \bibfield  {author} {\bibinfo {author} {\bibfnamefont {Y.-Z.}\ \bibnamefont
  {Chou}}, \bibinfo {author} {\bibfnamefont {Y.}~\bibnamefont {Liao}}, \ and\
  \bibinfo {author} {\bibfnamefont {M.~S.}\ \bibnamefont {Foster}},\ }\href
  {\doibase 10.1103/PhysRevB.95.104507} {\bibfield  {journal} {\bibinfo
  {journal} {Phys. Rev. B}\ }\textbf {\bibinfo {volume} {95}},\ \bibinfo
  {pages} {104507} (\bibinfo {year} {2017})}\BibitemShut {NoStop}%
\bibitem [{\citenamefont {Cui}\ \emph {et~al.}(2019)\citenamefont {Cui},
  \citenamefont {Sch\"utt}, \citenamefont {Orth},\ and\ \citenamefont
  {Fernandes}}]{Cui2019}%
  \BibitemOpen
  \bibfield  {author} {\bibinfo {author} {\bibfnamefont {T.}~\bibnamefont
  {Cui}}, \bibinfo {author} {\bibfnamefont {M.}~\bibnamefont {Sch\"utt}},
  \bibinfo {author} {\bibfnamefont {P.~P.}\ \bibnamefont {Orth}}, \ and\
  \bibinfo {author} {\bibfnamefont {R.~M.}\ \bibnamefont {Fernandes}},\ }\href
  {\doibase 10.1103/PhysRevB.100.144513} {\bibfield  {journal} {\bibinfo
  {journal} {Phys. Rev. B}\ }\textbf {\bibinfo {volume} {100}},\ \bibinfo
  {pages} {144513} (\bibinfo {year} {2019})}\BibitemShut {NoStop}%
\bibitem [{\citenamefont {Schwarz}\ \emph {et~al.}(2020)\citenamefont
  {Schwarz}, \citenamefont {Fauseweh}, \citenamefont {Tsuji}, \citenamefont
  {Cheng}, \citenamefont {Bittner}, \citenamefont {Krull}, \citenamefont
  {Berciu}, \citenamefont {Uhrig}, \citenamefont {Schnyder}, \citenamefont
  {Kaiser},\ and\ \citenamefont {Manske}}]{Schwarz2020}%
  \BibitemOpen
  \bibfield  {author} {\bibinfo {author} {\bibfnamefont {L.}~\bibnamefont
  {Schwarz}}, \bibinfo {author} {\bibfnamefont {B.}~\bibnamefont {Fauseweh}},
  \bibinfo {author} {\bibfnamefont {N.}~\bibnamefont {Tsuji}}, \bibinfo
  {author} {\bibfnamefont {N.}~\bibnamefont {Cheng}}, \bibinfo {author}
  {\bibfnamefont {N.}~\bibnamefont {Bittner}}, \bibinfo {author} {\bibfnamefont
  {H.}~\bibnamefont {Krull}}, \bibinfo {author} {\bibfnamefont
  {M.}~\bibnamefont {Berciu}}, \bibinfo {author} {\bibfnamefont {G.~S.}\
  \bibnamefont {Uhrig}}, \bibinfo {author} {\bibfnamefont {A.~P.}\ \bibnamefont
  {Schnyder}}, \bibinfo {author} {\bibfnamefont {S.}~\bibnamefont {Kaiser}}, \
  and\ \bibinfo {author} {\bibfnamefont {D.}~\bibnamefont {Manske}},\ }\href
  {\doibase 10.1038/s41467-019-13763-5} {\bibfield  {journal} {\bibinfo
  {journal} {Nature Communications}\ }\textbf {\bibinfo {volume} {11}},\
  \bibinfo {pages} {287} (\bibinfo {year} {2020})}\BibitemShut {NoStop}%
\bibitem [{\citenamefont {Mootz}\ \emph {et~al.}(2020)\citenamefont {Mootz},
  \citenamefont {Wang},\ and\ \citenamefont {Perakis}}]{Mootz2020}%
  \BibitemOpen
  \bibfield  {author} {\bibinfo {author} {\bibfnamefont {M.}~\bibnamefont
  {Mootz}}, \bibinfo {author} {\bibfnamefont {J.}~\bibnamefont {Wang}}, \ and\
  \bibinfo {author} {\bibfnamefont {I.~E.}\ \bibnamefont {Perakis}},\ }\href
  {\doibase 10.1103/PhysRevB.102.054517} {\bibfield  {journal} {\bibinfo
  {journal} {Phys. Rev. B}\ }\textbf {\bibinfo {volume} {102}},\ \bibinfo
  {pages} {054517} (\bibinfo {year} {2020})}\BibitemShut {NoStop}%
\bibitem [{\citenamefont {Matsunaga}\ \emph {et~al.}(2014)\citenamefont
  {Matsunaga}, \citenamefont {Tsuji}, \citenamefont {Fujita}, \citenamefont
  {Sugioka}, \citenamefont {Makise}, \citenamefont {Uzawa}, \citenamefont
  {Terai}, \citenamefont {Wang}, \citenamefont {Aoki},\ and\ \citenamefont
  {Shimano}}]{Matsunaga2014}%
  \BibitemOpen
  \bibfield  {author} {\bibinfo {author} {\bibfnamefont {R.}~\bibnamefont
  {Matsunaga}}, \bibinfo {author} {\bibfnamefont {N.}~\bibnamefont {Tsuji}},
  \bibinfo {author} {\bibfnamefont {H.}~\bibnamefont {Fujita}}, \bibinfo
  {author} {\bibfnamefont {A.}~\bibnamefont {Sugioka}}, \bibinfo {author}
  {\bibfnamefont {K.}~\bibnamefont {Makise}}, \bibinfo {author} {\bibfnamefont
  {Y.}~\bibnamefont {Uzawa}}, \bibinfo {author} {\bibfnamefont
  {H.}~\bibnamefont {Terai}}, \bibinfo {author} {\bibfnamefont
  {Z.}~\bibnamefont {Wang}}, \bibinfo {author} {\bibfnamefont {H.}~\bibnamefont
  {Aoki}}, \ and\ \bibinfo {author} {\bibfnamefont {R.}~\bibnamefont
  {Shimano}},\ }\href@noop {} {\bibfield  {journal} {\bibinfo  {journal}
  {Science}\ }\textbf {\bibinfo {volume} {345}},\ \bibinfo {pages} {1145}
  (\bibinfo {year} {2014})}\BibitemShut {NoStop}%
\bibitem [{\citenamefont {Tsuji}\ and\ \citenamefont {Aoki}(2015)}]{Tsuji2015}%
  \BibitemOpen
  \bibfield  {author} {\bibinfo {author} {\bibfnamefont {N.}~\bibnamefont
  {Tsuji}}\ and\ \bibinfo {author} {\bibfnamefont {H.}~\bibnamefont {Aoki}},\
  }\href@noop {} {\bibfield  {journal} {\bibinfo  {journal} {Phys. Rev. B}\
  }\textbf {\bibinfo {volume} {92}},\ \bibinfo {pages} {064508} (\bibinfo
  {year} {2015})}\BibitemShut {NoStop}%
\bibitem [{\citenamefont {Cea}\ \emph {et~al.}(2016)\citenamefont {Cea},
  \citenamefont {Castellani},\ and\ \citenamefont {Benfatto}}]{Cea2016}%
  \BibitemOpen
  \bibfield  {author} {\bibinfo {author} {\bibfnamefont {T.}~\bibnamefont
  {Cea}}, \bibinfo {author} {\bibfnamefont {C.}~\bibnamefont {Castellani}}, \
  and\ \bibinfo {author} {\bibfnamefont {L.}~\bibnamefont {Benfatto}},\ }\href
  {\doibase 10.1103/PhysRevB.93.180507} {\bibfield  {journal} {\bibinfo
  {journal} {Phys. Rev. B}\ }\textbf {\bibinfo {volume} {93}},\ \bibinfo
  {pages} {180507} (\bibinfo {year} {2016})}\BibitemShut {NoStop}%
\bibitem [{\citenamefont {Matsunaga}\ \emph {et~al.}(2017)\citenamefont
  {Matsunaga}, \citenamefont {Tsuji}, \citenamefont {Makise}, \citenamefont
  {Terai}, \citenamefont {Aoki},\ and\ \citenamefont
  {Shimano}}]{Matsunaga2017}%
  \BibitemOpen
  \bibfield  {author} {\bibinfo {author} {\bibfnamefont {R.}~\bibnamefont
  {Matsunaga}}, \bibinfo {author} {\bibfnamefont {N.}~\bibnamefont {Tsuji}},
  \bibinfo {author} {\bibfnamefont {K.}~\bibnamefont {Makise}}, \bibinfo
  {author} {\bibfnamefont {H.}~\bibnamefont {Terai}}, \bibinfo {author}
  {\bibfnamefont {H.}~\bibnamefont {Aoki}}, \ and\ \bibinfo {author}
  {\bibfnamefont {R.}~\bibnamefont {Shimano}},\ }\href {\doibase
  10.1103/PhysRevB.96.020505} {\bibfield  {journal} {\bibinfo  {journal} {Phys.
  Rev. B}\ }\textbf {\bibinfo {volume} {96}},\ \bibinfo {pages} {020505}
  (\bibinfo {year} {2017})}\BibitemShut {NoStop}%
\bibitem [{\citenamefont {Cea}\ \emph {et~al.}(2018)\citenamefont {Cea},
  \citenamefont {Barone}, \citenamefont {Castellani},\ and\ \citenamefont
  {Benfatto}}]{Cea2018}%
  \BibitemOpen
  \bibfield  {author} {\bibinfo {author} {\bibfnamefont {T.}~\bibnamefont
  {Cea}}, \bibinfo {author} {\bibfnamefont {P.}~\bibnamefont {Barone}},
  \bibinfo {author} {\bibfnamefont {C.}~\bibnamefont {Castellani}}, \ and\
  \bibinfo {author} {\bibfnamefont {L.}~\bibnamefont {Benfatto}},\ }\href
  {\doibase 10.1103/PhysRevB.97.094516} {\bibfield  {journal} {\bibinfo
  {journal} {Phys. Rev. B}\ }\textbf {\bibinfo {volume} {97}},\ \bibinfo
  {pages} {094516} (\bibinfo {year} {2018})}\BibitemShut {NoStop}%
\bibitem [{\citenamefont {Katsumi}\ \emph {et~al.}(2018)\citenamefont
  {Katsumi}, \citenamefont {Tsuji}, \citenamefont {Hamada}, \citenamefont
  {Matsunaga}, \citenamefont {Schneeloch}, \citenamefont {Zhong}, \citenamefont
  {Gu}, \citenamefont {Aoki}, \citenamefont {Gallais},\ and\ \citenamefont
  {Shimano}}]{Katsumi2018}%
  \BibitemOpen
  \bibfield  {author} {\bibinfo {author} {\bibfnamefont {K.}~\bibnamefont
  {Katsumi}}, \bibinfo {author} {\bibfnamefont {N.}~\bibnamefont {Tsuji}},
  \bibinfo {author} {\bibfnamefont {Y.~I.}\ \bibnamefont {Hamada}}, \bibinfo
  {author} {\bibfnamefont {R.}~\bibnamefont {Matsunaga}}, \bibinfo {author}
  {\bibfnamefont {J.}~\bibnamefont {Schneeloch}}, \bibinfo {author}
  {\bibfnamefont {R.~D.}\ \bibnamefont {Zhong}}, \bibinfo {author}
  {\bibfnamefont {G.~D.}\ \bibnamefont {Gu}}, \bibinfo {author} {\bibfnamefont
  {H.}~\bibnamefont {Aoki}}, \bibinfo {author} {\bibfnamefont {Y.}~\bibnamefont
  {Gallais}}, \ and\ \bibinfo {author} {\bibfnamefont {R.}~\bibnamefont
  {Shimano}},\ }\href {\doibase 10.1103/PhysRevLett.120.117001} {\bibfield
  {journal} {\bibinfo  {journal} {Phys. Rev. Lett.}\ }\textbf {\bibinfo
  {volume} {120}},\ \bibinfo {pages} {117001} (\bibinfo {year}
  {2018})}\BibitemShut {NoStop}%
\bibitem [{\citenamefont {Chu}\ \emph {et~al.}(2020)\citenamefont {Chu},
  \citenamefont {Kim}, \citenamefont {Katsumi}, \citenamefont {Kovalev},
  \citenamefont {Dawson}, \citenamefont {Schwarz}, \citenamefont {Yoshikawa},
  \citenamefont {Kim}, \citenamefont {Putzky}, \citenamefont {Li},
  \citenamefont {Raffy}, \citenamefont {Germanskiy}, \citenamefont {Deinert},
  \citenamefont {Awari}, \citenamefont {Ilyakov}, \citenamefont {Green},
  \citenamefont {Chen}, \citenamefont {Bawatna}, \citenamefont {Cristiani},
  \citenamefont {Logvenov}, \citenamefont {Gallais}, \citenamefont {Boris},
  \citenamefont {Keimer}, \citenamefont {Schnyder}, \citenamefont {Manske},
  \citenamefont {Gensch}, \citenamefont {Wang}, \citenamefont {Shimano},\ and\
  \citenamefont {Kaiser}}]{Chu2020}%
  \BibitemOpen
  \bibfield  {author} {\bibinfo {author} {\bibfnamefont {H.}~\bibnamefont
  {Chu}}, \bibinfo {author} {\bibfnamefont {M.-J.}\ \bibnamefont {Kim}},
  \bibinfo {author} {\bibfnamefont {K.}~\bibnamefont {Katsumi}}, \bibinfo
  {author} {\bibfnamefont {S.}~\bibnamefont {Kovalev}}, \bibinfo {author}
  {\bibfnamefont {R.~D.}\ \bibnamefont {Dawson}}, \bibinfo {author}
  {\bibfnamefont {L.}~\bibnamefont {Schwarz}}, \bibinfo {author} {\bibfnamefont
  {N.}~\bibnamefont {Yoshikawa}}, \bibinfo {author} {\bibfnamefont
  {G.}~\bibnamefont {Kim}}, \bibinfo {author} {\bibfnamefont {D.}~\bibnamefont
  {Putzky}}, \bibinfo {author} {\bibfnamefont {Z.~Z.}\ \bibnamefont {Li}},
  \bibinfo {author} {\bibfnamefont {H.}~\bibnamefont {Raffy}}, \bibinfo
  {author} {\bibfnamefont {S.}~\bibnamefont {Germanskiy}}, \bibinfo {author}
  {\bibfnamefont {J.-C.}\ \bibnamefont {Deinert}}, \bibinfo {author}
  {\bibfnamefont {N.}~\bibnamefont {Awari}}, \bibinfo {author} {\bibfnamefont
  {I.}~\bibnamefont {Ilyakov}}, \bibinfo {author} {\bibfnamefont
  {B.}~\bibnamefont {Green}}, \bibinfo {author} {\bibfnamefont
  {M.}~\bibnamefont {Chen}}, \bibinfo {author} {\bibfnamefont {M.}~\bibnamefont
  {Bawatna}}, \bibinfo {author} {\bibfnamefont {G.}~\bibnamefont {Cristiani}},
  \bibinfo {author} {\bibfnamefont {G.}~\bibnamefont {Logvenov}}, \bibinfo
  {author} {\bibfnamefont {Y.}~\bibnamefont {Gallais}}, \bibinfo {author}
  {\bibfnamefont {A.~V.}\ \bibnamefont {Boris}}, \bibinfo {author}
  {\bibfnamefont {B.}~\bibnamefont {Keimer}}, \bibinfo {author} {\bibfnamefont
  {A.~P.}\ \bibnamefont {Schnyder}}, \bibinfo {author} {\bibfnamefont
  {D.}~\bibnamefont {Manske}}, \bibinfo {author} {\bibfnamefont
  {M.}~\bibnamefont {Gensch}}, \bibinfo {author} {\bibfnamefont
  {Z.}~\bibnamefont {Wang}}, \bibinfo {author} {\bibfnamefont {R.}~\bibnamefont
  {Shimano}}, \ and\ \bibinfo {author} {\bibfnamefont {S.}~\bibnamefont
  {Kaiser}},\ }\href {\doibase 10.1038/s41467-020-15613-1} {\bibfield
  {journal} {\bibinfo  {journal} {Nature Communications}\ }\textbf {\bibinfo
  {volume} {11}},\ \bibinfo {pages} {1793} (\bibinfo {year}
  {2020})}\BibitemShut {NoStop}%
\bibitem [{\citenamefont {Jujo}(2018)}]{Jujo18}%
  \BibitemOpen
  \bibfield  {author} {\bibinfo {author} {\bibfnamefont {T.}~\bibnamefont
  {Jujo}},\ }\href {\doibase 10.7566/JPSJ.87.024704} {\bibfield  {journal}
  {\bibinfo  {journal} {Journal of the Physical Society of Japan}\ }\textbf
  {\bibinfo {volume} {87}},\ \bibinfo {pages} {024704} (\bibinfo {year}
  {2018})},\ \Eprint
  {http://arxiv.org/abs/https://doi.org/10.7566/JPSJ.87.024704}
  {https://doi.org/10.7566/JPSJ.87.024704} \BibitemShut {NoStop}%
\bibitem [{\citenamefont {Murotani}\ and\ \citenamefont
  {Shimano}(2019)}]{Murotani2019}%
  \BibitemOpen
  \bibfield  {author} {\bibinfo {author} {\bibfnamefont {Y.}~\bibnamefont
  {Murotani}}\ and\ \bibinfo {author} {\bibfnamefont {R.}~\bibnamefont
  {Shimano}},\ }\href {\doibase 10.1103/PhysRevB.99.224510} {\bibfield
  {journal} {\bibinfo  {journal} {Phys. Rev. B}\ }\textbf {\bibinfo {volume}
  {99}},\ \bibinfo {pages} {224510} (\bibinfo {year} {2019})}\BibitemShut
  {NoStop}%
\bibitem [{\citenamefont {Silaev}(2019)}]{Silaev2019}%
  \BibitemOpen
  \bibfield  {author} {\bibinfo {author} {\bibfnamefont {M.}~\bibnamefont
  {Silaev}},\ }\href {\doibase 10.1103/PhysRevB.99.224511} {\bibfield
  {journal} {\bibinfo  {journal} {Phys. Rev. B}\ }\textbf {\bibinfo {volume}
  {99}},\ \bibinfo {pages} {224511} (\bibinfo {year} {2019})}\BibitemShut
  {NoStop}%
\bibitem [{\citenamefont {Seibold}\ \emph {et~al.}(2021)\citenamefont
  {Seibold}, \citenamefont {Udina}, \citenamefont {Castellani},\ and\
  \citenamefont {Benfatto}}]{Seibold2021}%
  \BibitemOpen
  \bibfield  {author} {\bibinfo {author} {\bibfnamefont {G.}~\bibnamefont
  {Seibold}}, \bibinfo {author} {\bibfnamefont {M.}~\bibnamefont {Udina}},
  \bibinfo {author} {\bibfnamefont {C.}~\bibnamefont {Castellani}}, \ and\
  \bibinfo {author} {\bibfnamefont {L.}~\bibnamefont {Benfatto}},\ }\href
  {\doibase 10.1103/PhysRevB.103.014512} {\bibfield  {journal} {\bibinfo
  {journal} {Phys. Rev. B}\ }\textbf {\bibinfo {volume} {103}},\ \bibinfo
  {pages} {014512} (\bibinfo {year} {2021})}\BibitemShut {NoStop}%
\bibitem [{\citenamefont {Haenel}\ \emph {et~al.}(2021)\citenamefont {Haenel},
  \citenamefont {Froese}, \citenamefont {Manske},\ and\ \citenamefont
  {Schwarz}}]{Haenel2021}%
  \BibitemOpen
  \bibfield  {author} {\bibinfo {author} {\bibfnamefont {R.}~\bibnamefont
  {Haenel}}, \bibinfo {author} {\bibfnamefont {P.}~\bibnamefont {Froese}},
  \bibinfo {author} {\bibfnamefont {D.}~\bibnamefont {Manske}}, \ and\ \bibinfo
  {author} {\bibfnamefont {L.}~\bibnamefont {Schwarz}},\ }\href
  {https://arxiv.org/abs/2012.07674} {\bibfield  {journal} {\bibinfo  {journal}
  {arXiv:2012.07674}\ } (\bibinfo {year} {2021})}\BibitemShut {NoStop}%
\bibitem [{\citenamefont {Basov}\ and\ \citenamefont
  {Timusk}(2005)}]{Basov2005}%
  \BibitemOpen
  \bibfield  {author} {\bibinfo {author} {\bibfnamefont {D.~N.}\ \bibnamefont
  {Basov}}\ and\ \bibinfo {author} {\bibfnamefont {T.}~\bibnamefont {Timusk}},\
  }\href {\doibase 10.1103/RevModPhys.77.721} {\bibfield  {journal} {\bibinfo
  {journal} {Rev. Mod. Phys.}\ }\textbf {\bibinfo {volume} {77}},\ \bibinfo
  {pages} {721} (\bibinfo {year} {2005})}\BibitemShut {NoStop}%
\bibitem [{\citenamefont {Sun}\ \emph {et~al.}(2020)\citenamefont {Sun},
  \citenamefont {Fogler}, \citenamefont {Basov},\ and\ \citenamefont
  {Millis}}]{Sun2020}%
  \BibitemOpen
  \bibfield  {author} {\bibinfo {author} {\bibfnamefont {Z.}~\bibnamefont
  {Sun}}, \bibinfo {author} {\bibfnamefont {M.~M.}\ \bibnamefont {Fogler}},
  \bibinfo {author} {\bibfnamefont {D.~N.}\ \bibnamefont {Basov}}, \ and\
  \bibinfo {author} {\bibfnamefont {A.~J.}\ \bibnamefont {Millis}},\ }\href
  {\doibase 10.1103/PhysRevResearch.2.023413} {\bibfield  {journal} {\bibinfo
  {journal} {Phys. Rev. Research}\ }\textbf {\bibinfo {volume} {2}},\ \bibinfo
  {pages} {023413} (\bibinfo {year} {2020})}\BibitemShut {NoStop}%
\bibitem [{\citenamefont {Anderson}(1963)}]{Anderson1963}%
  \BibitemOpen
  \bibfield  {author} {\bibinfo {author} {\bibfnamefont {P.~W.}\ \bibnamefont
  {Anderson}},\ }\href {\doibase 10.1103/PhysRev.130.439} {\bibfield  {journal}
  {\bibinfo  {journal} {Phys. Rev.}\ }\textbf {\bibinfo {volume} {130}},\
  \bibinfo {pages} {439} (\bibinfo {year} {1963})}\BibitemShut {NoStop}%
\bibitem [{\citenamefont {Carlson}\ and\ \citenamefont
  {Goldman}(1975)}]{Carlson1975}%
  \BibitemOpen
  \bibfield  {author} {\bibinfo {author} {\bibfnamefont {R.~V.}\ \bibnamefont
  {Carlson}}\ and\ \bibinfo {author} {\bibfnamefont {A.~M.}\ \bibnamefont
  {Goldman}},\ }\href {\doibase 10.1103/PhysRevLett.34.11} {\bibfield
  {journal} {\bibinfo  {journal} {Phys. Rev. Lett.}\ }\textbf {\bibinfo
  {volume} {34}},\ \bibinfo {pages} {11} (\bibinfo {year} {1975})}\BibitemShut
  {NoStop}%
\bibitem [{\citenamefont {Bardasis}\ and\ \citenamefont
  {Schrieffer}(1961)}]{bardasis61}%
  \BibitemOpen
  \bibfield  {author} {\bibinfo {author} {\bibfnamefont {A.}~\bibnamefont
  {Bardasis}}\ and\ \bibinfo {author} {\bibfnamefont {J.~R.}\ \bibnamefont
  {Schrieffer}},\ }\href {\doibase 10.1103/PhysRev.121.1050} {\bibfield
  {journal} {\bibinfo  {journal} {Phys. Rev.}\ }\textbf {\bibinfo {volume}
  {121}},\ \bibinfo {pages} {1050} (\bibinfo {year} {1961})}\BibitemShut
  {NoStop}%
\bibitem [{\citenamefont {Kretzschmar}\ \emph {et~al.}(2013)\citenamefont
  {Kretzschmar}, \citenamefont {Muschler}, \citenamefont {B\"ohm},
  \citenamefont {Baum}, \citenamefont {Hackl}, \citenamefont {Wen},
  \citenamefont {Tsurkan}, \citenamefont {Deisenhofer},\ and\ \citenamefont
  {Loidl}}]{Kretschmar2013}%
  \BibitemOpen
  \bibfield  {author} {\bibinfo {author} {\bibfnamefont {F.}~\bibnamefont
  {Kretzschmar}}, \bibinfo {author} {\bibfnamefont {B.}~\bibnamefont
  {Muschler}}, \bibinfo {author} {\bibfnamefont {T.}~\bibnamefont {B\"ohm}},
  \bibinfo {author} {\bibfnamefont {A.}~\bibnamefont {Baum}}, \bibinfo {author}
  {\bibfnamefont {R.}~\bibnamefont {Hackl}}, \bibinfo {author} {\bibfnamefont
  {H.-H.}\ \bibnamefont {Wen}}, \bibinfo {author} {\bibfnamefont
  {V.}~\bibnamefont {Tsurkan}}, \bibinfo {author} {\bibfnamefont
  {J.}~\bibnamefont {Deisenhofer}}, \ and\ \bibinfo {author} {\bibfnamefont
  {A.}~\bibnamefont {Loidl}},\ }\href {\doibase 10.1103/PhysRevLett.110.187002}
  {\bibfield  {journal} {\bibinfo  {journal} {Phys. Rev. Lett.}\ }\textbf
  {\bibinfo {volume} {110}},\ \bibinfo {pages} {187002} (\bibinfo {year}
  {2013})}\BibitemShut {NoStop}%
\bibitem [{\citenamefont {B\"ohm}\ \emph {et~al.}(2014)\citenamefont {B\"ohm},
  \citenamefont {Kemper}, \citenamefont {Moritz}, \citenamefont {Kretzschmar},
  \citenamefont {Muschler}, \citenamefont {Eiter}, \citenamefont {Hackl},
  \citenamefont {Devereaux}, \citenamefont {Scalapino},\ and\ \citenamefont
  {Wen}}]{Boehm2014}%
  \BibitemOpen
  \bibfield  {author} {\bibinfo {author} {\bibfnamefont {T.}~\bibnamefont
  {B\"ohm}}, \bibinfo {author} {\bibfnamefont {A.~F.}\ \bibnamefont {Kemper}},
  \bibinfo {author} {\bibfnamefont {B.}~\bibnamefont {Moritz}}, \bibinfo
  {author} {\bibfnamefont {F.}~\bibnamefont {Kretzschmar}}, \bibinfo {author}
  {\bibfnamefont {B.}~\bibnamefont {Muschler}}, \bibinfo {author}
  {\bibfnamefont {H.-M.}\ \bibnamefont {Eiter}}, \bibinfo {author}
  {\bibfnamefont {R.}~\bibnamefont {Hackl}}, \bibinfo {author} {\bibfnamefont
  {T.~P.}\ \bibnamefont {Devereaux}}, \bibinfo {author} {\bibfnamefont {D.~J.}\
  \bibnamefont {Scalapino}}, \ and\ \bibinfo {author} {\bibfnamefont {H.-H.}\
  \bibnamefont {Wen}},\ }\href {\doibase 10.1103/PhysRevX.4.041046} {\bibfield
  {journal} {\bibinfo  {journal} {Phys. Rev. X}\ }\textbf {\bibinfo {volume}
  {4}},\ \bibinfo {pages} {041046} (\bibinfo {year} {2014})}\BibitemShut
  {NoStop}%
\bibitem [{\citenamefont {Wu}\ \emph {et~al.}(2017)\citenamefont {Wu},
  \citenamefont {Richard}, \citenamefont {Ding}, \citenamefont {Wen},
  \citenamefont {Tan}, \citenamefont {Wang}, \citenamefont {Zhang},
  \citenamefont {Dai},\ and\ \citenamefont {Blumberg}}]{wu17}%
  \BibitemOpen
  \bibfield  {author} {\bibinfo {author} {\bibfnamefont {S.-F.}\ \bibnamefont
  {Wu}}, \bibinfo {author} {\bibfnamefont {P.}~\bibnamefont {Richard}},
  \bibinfo {author} {\bibfnamefont {H.}~\bibnamefont {Ding}}, \bibinfo {author}
  {\bibfnamefont {H.-H.}\ \bibnamefont {Wen}}, \bibinfo {author} {\bibfnamefont
  {G.}~\bibnamefont {Tan}}, \bibinfo {author} {\bibfnamefont {M.}~\bibnamefont
  {Wang}}, \bibinfo {author} {\bibfnamefont {C.}~\bibnamefont {Zhang}},
  \bibinfo {author} {\bibfnamefont {P.}~\bibnamefont {Dai}}, \ and\ \bibinfo
  {author} {\bibfnamefont {G.}~\bibnamefont {Blumberg}},\ }\href {\doibase
  10.1103/PhysRevB.95.085125} {\bibfield  {journal} {\bibinfo  {journal} {Phys.
  Rev. B}\ }\textbf {\bibinfo {volume} {95}},\ \bibinfo {pages} {085125}
  (\bibinfo {year} {2017})}\BibitemShut {NoStop}%
\bibitem [{\citenamefont {B\"ohm}\ \emph {et~al.}(2018)\citenamefont {B\"ohm},
  \citenamefont {Kretzschmar}, \citenamefont {Baum}, \citenamefont {Rehm},
  \citenamefont {Jost}, \citenamefont {Ahangharnejhad}, \citenamefont
  {Thomale}, \citenamefont {Platt}, \citenamefont {Maier}, \citenamefont
  {Hanke}, \citenamefont {Moritz}, \citenamefont {Devereaux}, \citenamefont
  {Scalapino}, \citenamefont {Maiti}, \citenamefont {Hirschfeld}, \citenamefont
  {Adelmann}, \citenamefont {Wolf}, \citenamefont {Wen},\ and\ \citenamefont
  {Hackl}}]{Boehm18}%
  \BibitemOpen
  \bibfield  {author} {\bibinfo {author} {\bibfnamefont {T.}~\bibnamefont
  {B\"ohm}}, \bibinfo {author} {\bibfnamefont {F.}~\bibnamefont {Kretzschmar}},
  \bibinfo {author} {\bibfnamefont {A.}~\bibnamefont {Baum}}, \bibinfo {author}
  {\bibfnamefont {M.}~\bibnamefont {Rehm}}, \bibinfo {author} {\bibfnamefont
  {D.}~\bibnamefont {Jost}}, \bibinfo {author} {\bibfnamefont {R.~H.}\
  \bibnamefont {Ahangharnejhad}}, \bibinfo {author} {\bibfnamefont
  {R.}~\bibnamefont {Thomale}}, \bibinfo {author} {\bibfnamefont
  {C.}~\bibnamefont {Platt}}, \bibinfo {author} {\bibfnamefont {T.~A.}\
  \bibnamefont {Maier}}, \bibinfo {author} {\bibfnamefont {W.}~\bibnamefont
  {Hanke}}, \bibinfo {author} {\bibfnamefont {B.}~\bibnamefont {Moritz}},
  \bibinfo {author} {\bibfnamefont {T.~P.}\ \bibnamefont {Devereaux}}, \bibinfo
  {author} {\bibfnamefont {D.~J.}\ \bibnamefont {Scalapino}}, \bibinfo {author}
  {\bibfnamefont {S.}~\bibnamefont {Maiti}}, \bibinfo {author} {\bibfnamefont
  {P.~J.}\ \bibnamefont {Hirschfeld}}, \bibinfo {author} {\bibfnamefont
  {P.}~\bibnamefont {Adelmann}}, \bibinfo {author} {\bibfnamefont
  {T.}~\bibnamefont {Wolf}}, \bibinfo {author} {\bibfnamefont {H.-H.}\
  \bibnamefont {Wen}}, \ and\ \bibinfo {author} {\bibfnamefont
  {R.}~\bibnamefont {Hackl}},\ }\href {\doibase
  doi.org/10.1038/s41535-018-0118-z} {\bibfield  {journal} {\bibinfo  {journal}
  {npj Quantum Materials}\ }\textbf {\bibinfo {volume} {3}},\ \bibinfo {pages}
  {48} (\bibinfo {year} {2018})}\BibitemShut {NoStop}%
\bibitem [{\citenamefont {Jost}\ \emph {et~al.}(2018)\citenamefont {Jost},
  \citenamefont {Scholz}, \citenamefont {Zweck}, \citenamefont {Meier},
  \citenamefont {B\"ohmer}, \citenamefont {Canfield}, \citenamefont
  {Lazarevi\ifmmode~\acute{c}\else \'{c}\fi{}},\ and\ \citenamefont
  {Hackl}}]{Jost2018}%
  \BibitemOpen
  \bibfield  {author} {\bibinfo {author} {\bibfnamefont {D.}~\bibnamefont
  {Jost}}, \bibinfo {author} {\bibfnamefont {J.-R.}\ \bibnamefont {Scholz}},
  \bibinfo {author} {\bibfnamefont {U.}~\bibnamefont {Zweck}}, \bibinfo
  {author} {\bibfnamefont {W.~R.}\ \bibnamefont {Meier}}, \bibinfo {author}
  {\bibfnamefont {A.~E.}\ \bibnamefont {B\"ohmer}}, \bibinfo {author}
  {\bibfnamefont {P.~C.}\ \bibnamefont {Canfield}}, \bibinfo {author}
  {\bibfnamefont {N.}~\bibnamefont {Lazarevi\ifmmode~\acute{c}\else
  \'{c}\fi{}}}, \ and\ \bibinfo {author} {\bibfnamefont {R.}~\bibnamefont
  {Hackl}},\ }\href {\doibase 10.1103/PhysRevB.98.020504} {\bibfield  {journal}
  {\bibinfo  {journal} {Phys. Rev. B}\ }\textbf {\bibinfo {volume} {98}},\
  \bibinfo {pages} {020504} (\bibinfo {year} {2018})}\BibitemShut {NoStop}%
\bibitem [{\citenamefont {He}\ \emph {et~al.}(2020)\citenamefont {He},
  \citenamefont {Li}, \citenamefont {Jost}, \citenamefont {Baum}, \citenamefont
  {Shen}, \citenamefont {Dong}, \citenamefont {Zhao},\ and\ \citenamefont
  {Hackl}}]{He2020}%
  \BibitemOpen
  \bibfield  {author} {\bibinfo {author} {\bibfnamefont {G.}~\bibnamefont
  {He}}, \bibinfo {author} {\bibfnamefont {D.}~\bibnamefont {Li}}, \bibinfo
  {author} {\bibfnamefont {D.}~\bibnamefont {Jost}}, \bibinfo {author}
  {\bibfnamefont {A.}~\bibnamefont {Baum}}, \bibinfo {author} {\bibfnamefont
  {P.~P.}\ \bibnamefont {Shen}}, \bibinfo {author} {\bibfnamefont {X.~L.}\
  \bibnamefont {Dong}}, \bibinfo {author} {\bibfnamefont {Z.~X.}\ \bibnamefont
  {Zhao}}, \ and\ \bibinfo {author} {\bibfnamefont {R.}~\bibnamefont {Hackl}},\
  }\href {\doibase 10.1103/PhysRevLett.125.217002} {\bibfield  {journal}
  {\bibinfo  {journal} {Phys. Rev. Lett.}\ }\textbf {\bibinfo {volume} {125}},\
  \bibinfo {pages} {217002} (\bibinfo {year} {2020})}\BibitemShut {NoStop}%
\bibitem [{\citenamefont {Maiti}\ and\ \citenamefont
  {Hirschfeld}(2015)}]{maiti15}%
  \BibitemOpen
  \bibfield  {author} {\bibinfo {author} {\bibfnamefont {S.}~\bibnamefont
  {Maiti}}\ and\ \bibinfo {author} {\bibfnamefont {P.~J.}\ \bibnamefont
  {Hirschfeld}},\ }\href {\doibase 10.1103/PhysRevB.92.094506} {\bibfield
  {journal} {\bibinfo  {journal} {Phys. Rev. B}\ }\textbf {\bibinfo {volume}
  {92}},\ \bibinfo {pages} {094506} (\bibinfo {year} {2015})}\BibitemShut
  {NoStop}%
\bibitem [{\citenamefont {Maiti}\ \emph {et~al.}(2016)\citenamefont {Maiti},
  \citenamefont {Maier}, \citenamefont {B\"ohm}, \citenamefont {Hackl},\ and\
  \citenamefont {Hirschfeld}}]{maiti16}%
  \BibitemOpen
  \bibfield  {author} {\bibinfo {author} {\bibfnamefont {S.}~\bibnamefont
  {Maiti}}, \bibinfo {author} {\bibfnamefont {T.~A.}\ \bibnamefont {Maier}},
  \bibinfo {author} {\bibfnamefont {T.}~\bibnamefont {B\"ohm}}, \bibinfo
  {author} {\bibfnamefont {R.}~\bibnamefont {Hackl}}, \ and\ \bibinfo {author}
  {\bibfnamefont {P.~J.}\ \bibnamefont {Hirschfeld}},\ }\href {\doibase
  10.1103/PhysRevLett.117.257001} {\bibfield  {journal} {\bibinfo  {journal}
  {Phys. Rev. Lett.}\ }\textbf {\bibinfo {volume} {117}},\ \bibinfo {pages}
  {257001} (\bibinfo {year} {2016})}\BibitemShut {NoStop}%
\bibitem [{\citenamefont {Allocca}\ \emph {et~al.}(2019)\citenamefont
  {Allocca}, \citenamefont {Raines}, \citenamefont {Curtis},\ and\
  \citenamefont {Galitski}}]{Allocca2019}%
  \BibitemOpen
  \bibfield  {author} {\bibinfo {author} {\bibfnamefont {A.~A.}\ \bibnamefont
  {Allocca}}, \bibinfo {author} {\bibfnamefont {Z.~M.}\ \bibnamefont {Raines}},
  \bibinfo {author} {\bibfnamefont {J.~B.}\ \bibnamefont {Curtis}}, \ and\
  \bibinfo {author} {\bibfnamefont {V.~M.}\ \bibnamefont {Galitski}},\ }\href
  {\doibase 10.1103/PhysRevB.99.020504} {\bibfield  {journal} {\bibinfo
  {journal} {Phys. Rev. B}\ }\textbf {\bibinfo {volume} {99}},\ \bibinfo
  {pages} {020504} (\bibinfo {year} {2019})}\BibitemShut {NoStop}%
\bibitem [{\citenamefont {M\"uller}\ \emph {et~al.}(2018)\citenamefont
  {M\"uller}, \citenamefont {Shen}, \citenamefont {Dzero},\ and\ \citenamefont
  {Eremin}}]{mueller18}%
  \BibitemOpen
  \bibfield  {author} {\bibinfo {author} {\bibfnamefont {M.~A.}\ \bibnamefont
  {M\"uller}}, \bibinfo {author} {\bibfnamefont {P.}~\bibnamefont {Shen}},
  \bibinfo {author} {\bibfnamefont {M.}~\bibnamefont {Dzero}}, \ and\ \bibinfo
  {author} {\bibfnamefont {I.}~\bibnamefont {Eremin}},\ }\href {\doibase
  10.1103/PhysRevB.98.024522} {\bibfield  {journal} {\bibinfo  {journal} {Phys.
  Rev. B}\ }\textbf {\bibinfo {volume} {98}},\ \bibinfo {pages} {024522}
  (\bibinfo {year} {2018})}\BibitemShut {NoStop}%
\bibitem [{\citenamefont {M\"uller}\ \emph {et~al.}(2019)\citenamefont
  {M\"uller}, \citenamefont {Volkov}, \citenamefont {Paul},\ and\ \citenamefont
  {Eremin}}]{Mueller19}%
  \BibitemOpen
  \bibfield  {author} {\bibinfo {author} {\bibfnamefont {M.~A.}\ \bibnamefont
  {M\"uller}}, \bibinfo {author} {\bibfnamefont {P.~A.}\ \bibnamefont
  {Volkov}}, \bibinfo {author} {\bibfnamefont {I.}~\bibnamefont {Paul}}, \ and\
  \bibinfo {author} {\bibfnamefont {I.~M.}\ \bibnamefont {Eremin}},\ }\href
  {\doibase 10.1103/PhysRevB.100.140501} {\bibfield  {journal} {\bibinfo
  {journal} {Phys. Rev. B}\ }\textbf {\bibinfo {volume} {100}},\ \bibinfo
  {pages} {140501} (\bibinfo {year} {2019})}\BibitemShut {NoStop}%
\bibitem [{\citenamefont {M\"uller}\ \emph {et~al.}(2021)\citenamefont
  {M\"uller}, \citenamefont {Volkov}, \citenamefont {Paul},\ and\ \citenamefont
  {Eremin}}]{Mueller21}%
  \BibitemOpen
  \bibfield  {author} {\bibinfo {author} {\bibfnamefont {M.~A.}\ \bibnamefont
  {M\"uller}}, \bibinfo {author} {\bibfnamefont {P.~A.}\ \bibnamefont
  {Volkov}}, \bibinfo {author} {\bibfnamefont {I.}~\bibnamefont {Paul}}, \ and\
  \bibinfo {author} {\bibfnamefont {I.~M.}\ \bibnamefont {Eremin}},\ }\href
  {\doibase 10.1103/PhysRevB.103.024519} {\bibfield  {journal} {\bibinfo
  {journal} {Phys. Rev. B}\ }\textbf {\bibinfo {volume} {103}},\ \bibinfo
  {pages} {024519} (\bibinfo {year} {2021})}\BibitemShut {NoStop}%
\bibitem [{\citenamefont {Paramekanti}\ \emph {et~al.}(2000)\citenamefont
  {Paramekanti}, \citenamefont {Randeria}, \citenamefont {Ramakrishnan},\ and\
  \citenamefont {Mandal}}]{Paramekanti2000}%
  \BibitemOpen
  \bibfield  {author} {\bibinfo {author} {\bibfnamefont {A.}~\bibnamefont
  {Paramekanti}}, \bibinfo {author} {\bibfnamefont {M.}~\bibnamefont
  {Randeria}}, \bibinfo {author} {\bibfnamefont {T.~V.}\ \bibnamefont
  {Ramakrishnan}}, \ and\ \bibinfo {author} {\bibfnamefont {S.~S.}\
  \bibnamefont {Mandal}},\ }\href {\doibase 10.1103/PhysRevB.62.6786}
  {\bibfield  {journal} {\bibinfo  {journal} {Phys. Rev. B}\ }\textbf {\bibinfo
  {volume} {62}},\ \bibinfo {pages} {6786} (\bibinfo {year}
  {2000})}\BibitemShut {NoStop}%
\bibitem [{\citenamefont {Benfatto}\ \emph {et~al.}(2004)\citenamefont
  {Benfatto}, \citenamefont {Toschi},\ and\ \citenamefont
  {Caprara}}]{Benfatto2004}%
  \BibitemOpen
  \bibfield  {author} {\bibinfo {author} {\bibfnamefont {L.}~\bibnamefont
  {Benfatto}}, \bibinfo {author} {\bibfnamefont {A.}~\bibnamefont {Toschi}}, \
  and\ \bibinfo {author} {\bibfnamefont {S.}~\bibnamefont {Caprara}},\ }\href
  {\doibase 10.1103/PhysRevB.69.184510} {\bibfield  {journal} {\bibinfo
  {journal} {Phys. Rev. B}\ }\textbf {\bibinfo {volume} {69}},\ \bibinfo
  {pages} {184510} (\bibinfo {year} {2004})}\BibitemShut {NoStop}%
\bibitem [{\citenamefont {Cea}\ \emph {et~al.}(2015)\citenamefont {Cea},
  \citenamefont {Castellani}, \citenamefont {Seibold},\ and\ \citenamefont
  {Benfatto}}]{Cea2015}%
  \BibitemOpen
  \bibfield  {author} {\bibinfo {author} {\bibfnamefont {T.}~\bibnamefont
  {Cea}}, \bibinfo {author} {\bibfnamefont {C.}~\bibnamefont {Castellani}},
  \bibinfo {author} {\bibfnamefont {G.}~\bibnamefont {Seibold}}, \ and\
  \bibinfo {author} {\bibfnamefont {L.}~\bibnamefont {Benfatto}},\ }\href
  {\doibase 10.1103/PhysRevLett.115.157002} {\bibfield  {journal} {\bibinfo
  {journal} {Phys. Rev. Lett.}\ }\textbf {\bibinfo {volume} {115}},\ \bibinfo
  {pages} {157002} (\bibinfo {year} {2015})}\BibitemShut {NoStop}%
\end{thebibliography}%
\newpage

\widetext

\begin{appendix}


\section{Derivation of effective action\label{appendix:a}}
In this section we introduce the model for our superconducting system with competing instabilities and derive an effective action in terms of the gaussian fluctuations following Refs. \cite{Paramekanti2000,Benfatto2004}. The full action is given by the kinetic part, the superconducting interaction and the long-range Coulomb interaction
\begin{align}
	S = S_{0} + S_\text{sc} + S_\text{c}.
\end{align}
The kinetic part takes nearest neighbor hopping on a square lattice into account and reads
\begin{align}\label{eq:S0}
S_0 = \int d\tau \left[ \sum_{\rr\sigma} c_{\rr\sigma}^\dagger(\tau)(\partial_\tau - \mu)c_{\rr\sigma}(\tau) - t\sum_{\langle\rr,\rr^\prime\rangle}c_{\rr\sigma}^\dagger(\tau) c_{\rr^\prime\sigma}(\tau)\right],
\end{align}
where $\mu$ is the chemical potential and $t$ is the hopping parameter. We choose the chemical potential such that the band filling is far from half-filling and the Fermi surface is near circular.
The superconducting interaction is assumed to have the form
\begin{align}\label{eq:H_sc}
    S_{\text{sc}}= -\int d\tau \sum_{\kk,\kk^\prime,\kq} \left(V_s + V_d\gamma_{\kk,d}\gamma_{\kk^\prime,d}\right)B^\dagger_{\kk,\kq}(\tau) B_{\kk^\prime,\kq}(\tau)  ,
\end{align}
with the short-hand notation $B_{\kk,\kq}(\tau)=  c_{-\kk+\kq/2,\downarrow}(\tau)c_{\kk+\kq/2,\uparrow}(\tau)$ for the spin singlet Cooper pairing. We choose the signs of the $s$- and $d$-wave interactions $V_s$ and $V_d$ such that $V_{s/d}>0$ implies attractive interaction. This four-fermion interaction is now decoupled using a standard Hubbard-Stratonovic transformation in the $s$-wave pairing channel $\sum_\kk B_{\kk,\kq}(\tau)$ and the $d$-wave pairing channel $\sum_\kk \gd B_{\kk,\kq}(\tau)$ and one obtains
\begin{align}
	S_\text{sc} = \int d\tau \left[  \sum_{\kq}\left(\frac{1}{V_s}\Delta^\dagger_{s}(\kq,\tau)\Delta_{s}(\kq,\tau) + \frac{1}{V_d}\Delta^\dagger_{d}(\kq,\tau)\Delta_{d}(\kq,\tau)\right) + \sum_{\kk,\kq}\left(\Delta_\kk(\kq,\tau)B^\dagger_{\kk,\kq} + \Delta^\dagger_\kk(\kq,\tau)B_{\kk,\kq}\right)\right],
\end{align}
where $\Delta_\kk(\kq,\tau) = \Delta_s(\kq\tau) + \Delta_d(\kq,\tau)\gd$ contains the two introduced Hubbard-Stratonovic fields $\Delta_s(\kq,\tau)$ and $\Delta_d(\kq,\tau)$, which transform according to the $A_{1g}$ ($s$-wave) and $B_{1g}$ ($d_{x^2-y^2}$-wave) irreducible representation of the tetragonal lattice. We choose the phase of $\Delta_s(\kq,\tau)$ to be real and positive by performing a gauge transformation for the global phase $c_{\rr\sigma}(\tau) \rightarrow c_{\rr\sigma}(\tau) e^{i \theta(\rr,\tau)/2}$. Here we assume that the phase does not change too fast as a function of lattice site. We can transform the action $S_{\text{sc}}$ from imaginary time to Matsubara frequency description and split the fields into saddle point value at $q \equiv (\kq,\vm) = 0$ plus fluctuations $\Delta_{s,d}^0 + \Delta_{s,d}(q)$ to obtain
\begin{align}\label{eq:eff_action_sc}
	S_\text{sc} = \sum_{k,k^\prime}\Psi_{k}&\Bigg[(\Delta_{s,0} + \Delta^\prime_{d,0}\gd(\kk))\delta_{k,k^\prime}\sigma_1 + \Delta_{d,0}^{\prime\prime}\gd(\kk^\prime)\delta_{k,k^\prime}\sigma_2 \nn \\
 +&\Sigma_{\Delta_s}(k,k^\prime) + \Sigma_{\Delta_d^\prime}(k,k^\prime) + \Sigma_{\Delta_d^{\prime\prime}}(k,k^\prime) \Bigg]\Psi_{k^\prime},
\end{align}
where we introduced the Nambu-Spinor $\Psi_k^\dagger = \left(c_{\kk,n}^\dagger, -c_{-\kk,-n}^\dagger\right)$ with $k = (\kk,\wn)$. Additionally we split the complex $d$-wave field $\Delta_d = \Delta_d^\prime -i\Delta_d^{\prime\prime}$ into two real fields. While the first line of Eq. \ref{eq:eff_action_sc} corresponds to the saddle point action, the second one describes the self-energy corrections due to fluctuations around the saddle point.
\begin{align}
	\Sigma_{\Delta_s}(k,k^\prime) &= \Delta_s(k-k^\prime)\sigma_1 \\
	\Sigma_{\Delta_d^\prime}(k,k^\prime) &= \Delta_d^{\prime}(k-k^\prime)\gamma_d((\kk+\kk^\prime)/2)\sigma_1 \\
	\Sigma_{\Delta_d^{\prime\prime}}(k,k^\prime) &= \Delta_d^{\prime\prime}(k-k^\prime)\gamma_d((\kk+\kk^\prime)/2)\sigma_2.
\end{align}
Here, we introduced the short-hand notation $k = (\kk,\wn)$
By performing the gauge transformation, the fluctuations of the global phase $\theta$ contribute to the kinetic action $S_{0}$ in Eq. (\ref{eq:S0}). After performing a Fourier transformation they read
\begin{align}
	S_0 = \sum_{k}\Psi^\dagger_{\kk n}&\Big[\left(  -\wn\sigma_0 + \xi_\kk \sigma_3 - \mu\sigma_3 \right)\delta_{k,k^\prime}\\ &+ \Sigma_{\theta_1}(k,k^\prime) + \Sigma_{\theta_2}(k,k^\prime) + \Sigma_{\theta_3}(k,k^\prime) \Big]\Psi_{k^\prime}
\end{align}
The remaining self-energy contributions are given by
\begin{align}
	\Sigma_{\theta_1}(k,k^\prime)  =& - \frac{i}{2}\left(i\nu_{n-n^\prime}\right)\theta(k-k^\prime)\sigma_3  \\	
	\Sigma_{\theta_2}(k,k^\prime)  =& 
	\frac{i}{2}\left(\xi_\kk - \xi_{\kk^\prime}\right)\theta(k-k^\prime)\sigma_0 
    \\
	\Sigma_{\theta_3}(k,k^\prime)  =& 	\frac{1}{2}\sum_{\substack{q_1,q_2,i } }\theta(q_1)\theta(q_2)\sin(\frac{\kq_{1,i}}{2})\sin(\frac{\kq_{2,i}}{2})\eval{\frac{\partial^2\xi_{\tilde{\kk}}}{\partial \tilde{k}_i^2}}_{\tilde{\kk} = \frac{\kk + \kk^\prime}{2}} \nn\\
	&\cdot\delta(q_1 + q_2 -(k-k^\prime))\label{eq:sigmatheta3}
\end{align}
To ensure correct renormalization of the phase fluctuations we include the effect of the long-range Coulomb interaction to our system and add the action $S_\text{C}$ to the system
\begin{align}
	S_{c} = \int d\tau\sum_{\substack{\kk,\kk^\prime,\kq\\\sigma,\sigma^\prime}}\frac{1}{2}V_{\kq}c^\dagger_{\kk+\kq,\sigma}(\tau)c^\dagger_{\kk^\prime-\kq,\sigma^\prime}(\tau)c_{\kk^\prime,\sigma^\prime}(\tau)c_{\kk,\sigma}(\tau),
\end{align}
where $V_{\kq} = 2\pi e^2/|\kq|$ is the Coulomb interaction between quasiparticles projected onto the 2D lattice. Performing a Hubbard-Stratonovic transformation introduces the field $\rho(q)$, and one obtains the self-energy due to the charged field
\begin{align}
	S_{c} = & \sum_q\left[-\frac{2}{V_{\kq}}\rho(-q)\rho(q) + \Psi_{k}\rho(k-k^\prime)\sigma_3\Psi_{k^\prime}\right] \nn\\
	=& \sum_q\left[-\frac{2}{V_{\kq}}\rho(-q)\rho(q) + \Psi^\dagger_{k}\Sigma_\rho(k,k^\prime)\Psi_{k^\prime}\right] 
\end{align}
with the self-energy contribution
\begin{align}\label{eq:sigma_rho}
	\Sigma_\rho(k,k^\prime) = \rho(k-k^\prime)\sigma_3.
\end{align}
Finally, we include the effect of an applied pulsed electric field by introducing a time-dependent vector potential $\mathbf{A}(t)$. This couples to the electric field via the Peierls substitution $c^\dagger_{\rr,\sigma}c_{\rr+\boldsymbol{\delta},\sigma}\rightarrow e^{ie\mathbf{A}\cdot\boldsymbol{\delta}/c}c^\dagger_{\rr,\sigma}c_{\rr+\boldsymbol{\delta},\sigma}$, which translates into a shift for the dispersion $\xi_\kk \rightarrow \xi_{\kk-\frac{e}{c}\mathbf{A}}$. Note that we neglect spacial variation of the vector potential, which implies that it carries zero kinetic momentum transfer $\kq=0$. This is justified, because the wavelength of the light used in THz experiments is much larger than a typical coherence length in unconventional superconductors. Expanding the dispersion up to a second order in the vector potential $\xi_{\kk-\frac{e}{c}\mathbf{A}} =\simeq \xi_\kk -\frac{e}{c}A_i(t)\frac{\partial \xi_\kk}{\partial k_i} +\frac{e^2}{c^2}A^2_i(t)\frac{\partial^2 \xi_\kk}{\partial^2 k_i}$ one obtains the corrections 
\begin{align}
	\Sigma_{A_i}(k,k^\prime) &= -e\sum_i A_i(\wn - \wn^\prime)\frac{\partial \xi_\kk}{\partial k_i} \delta_{\kk,\kk^\prime}\sigma_0 \label{eq:A} \\
	\Sigma_{A^2_i}(k,k^\prime) &=  \frac{e^2}{2}\sum_i A_i^2(\wn - \wn^\prime)\frac{\partial^2 \xi_\kk}{\partial k_i^2} \delta_{\kk,\kk^\prime}\sigma_3 \label{eq:A2}
\end{align}
Since the third-harmonic generated current stems from contributions to the effective action, which are quartic in the vector potential $\mathbf{A}$ the self-energy correction in eq. \ref{eq:A} does not contribute to the third-harmonic generated current up to quadratic order. Therefore we only focus on the contribution given by Eq. \ref{eq:A2}.\\
The total action now reads
\begin{align}
	S = \sum_{k,k^\prime}\Psi^\dagger_{k}\Big[-G_{0}^{-1}(k)\delta_{k,k^\prime}+ \Sigma(k,k^\prime)\Big]\Psi_{k},
\end{align}
with the saddle point Greens function $G_0(k) = \left(\wn\sigma_0 - \xi\sigma_3  -\Delta\sigma_1 \right)^{-1}$ and the combined self-energy contributions
\begin{align}
    \Sigma(k,k^\prime) = \Sigma_{\Delta_s}(k,k^\prime) + \Sigma_{\Delta_d^\prime}(k,k^\prime) + \Sigma_{\Delta_d^{\prime\prime}}(k,k^\prime) +  \Sigma_{\theta_1}(k,k^\prime) + \Sigma_{\theta_2}(k,k^\prime) + \Sigma_{\theta_3}(k,k^\prime) + \Sigma_\rho(k,k^\prime) + \Sigma_{A_i^2}(k,k^\prime).
\end{align}
Integrating out the fermions and expanding the action for small fluctuations around the saddle point yields 
\begin{align}
S_{fl} =  \frac{1}{2}\sum_{i,j}A_i^2(-\vm)K_{0,ij}(\vm)A_j^2(\vm)+\sum_\alpha\eta_{\alpha}^T(-\vm)\chi_{\eta_\alpha,A^2_i}(\vm)A^2_i(\vm)+ \frac{1}{2} \boldsymbol{\eta}^T(-q) \hat{\chi}(q)\boldsymbol{\eta}(q) ,
\end{align}
where the bare current-current kernel $K_{0,ij}$ is defined via
\begin{align}\label{eq:K0}
   K_{0,ij} & = \sum_\kk\sum_\wn \frac{\partial^2\xi_\kk}{\partial k_i ^2}\frac{\partial^2\xi_\kk}{\partial k_j ^2}\tr(G_0(\kk,\wn+\vm)\sigma_3 G_0(\kk,\wn)\sigma_3)\nn \\
&=-\sum_\kk   4\Delta^2 \frac{\partial^2\xi_\kk}{\partial k_i ^2}\frac{\partial^2\xi_\kk}{\partial k_j ^2}F_\kk(\vm) 
\end{align}
with the function
\begin{align}
    F_\kk(\vm) = \frac{\tanh(\beta E_\kk /2)}{4E_\kk (E_\kk^2 - (\vm)^2)}
\end{align}
vector  $\boldsymbol{\eta}(q) = \left(\Delta_s(q),\theta(q),\Delta_d^\prime(q),\Delta_d^{\prime\prime}(q),\rho(q)\right)^T$ contains the fields and their fluctuations on a Gaussian level is given by a matrix $\hat{\chi}$ whose diagonal elements $\chi_{\eta_{\alpha}\eta_{\beta}}$ read
\begin{align}
	\chi_{\Delta_{s}\Delta_{s}} = \frac{2}{V_s} + \sum_{\kk,\wn}\tr(G_0(\kk,\wn+i\nu_m)\sigma_1G_0(\kk,\wn)\sigma_1)
	=\sum_\kk \left(4\Delta^2_\kk-(\vm)^2\right) F_\kk(\vm)
\end{align}
\begin{align}
	\chi_{\Delta_d^\prime\Delta_d^\prime}(i\nu_m) =\frac{2}{V_d} + \sum_{\kk,\wn}\gd^2\tr(G_0(\kk,\wn+i\nu_m)\sigma_1G_0(\kk,\wn)\sigma_1) = \frac{2}{V_d} - \sum_\kk (4\xi_\kk\gd^2)F_\kk(\vm).
\end{align}
\begin{align}
 \chi_{\Delta^{\prime\prime}_d\Delta^{\prime\prime}_d}(i\nu_m)=\frac{2}{V_d}+\sum_{\kk,\wn}\gd^2\tr(G_0(\kk,\wn+i\nu_m)\sigma_2G_0(\kk,\wn)\sigma_2) 
=\frac{2}{V_d} -\sum_\kk \left( -4E^2_\kk \gd^2 \right) F_\kk(\vm)
\end{align}
\begin{align}
    \chi_{\theta\theta}(\vm) &= \frac{(\vm)^2}{4}\sum_\kk\sum_\wn \tr(G_0(\kk,\wn+\vm)\sigma_3G_0(\kk,\wn)\sigma_3) +  \frac{1}{4}n_{s}\kq^2 \nn\\&= -(\vm)^2\sum_\kk \Delta^2F_\kk(\vm) +  \frac{1}{4}n_{s}\kq^2.\label{eq:chi_thetatheta}
\end{align}
\begin{align}
    \tilde{\chi}_{\rho\rho}(\vm) &= -\frac{1}{V_\kq} + \sum_\kk\sum_\wn \tr(G_0(\kk,\wn+\vm)\sigma_3G_0(\kk,\wn)\sigma_3) =-\frac{1}{V_\kq} -\sum_\kk   4\Delta^2(\kk)F_\kk(\vm)
\end{align}
For Eq. \ref{eq:chi_thetatheta} we introduced the superfluid stiffness $n_s = \sum_\kk \frac{\partial^2 \xi_\kk}{\partial k_i^2}\left(\frac{E_\kk -\xi_\kk\tanh{\beta E_\kk/2}}{E_\kk}\right)$. The off-diagonal non-zero terms are
\begin{align}\label{eq:chi_ts_0}
\chi_{\theta\Delta_s}(\vm) &= i\frac{(\vm)}{2}\sum_\kk\sum_\wn \tr(G_0(\kk,\wn+\vm)\sigma_1G_0(\kk,\wn)\sigma_3) \\
&=2i(\vm)\sum_\kk \xi_\kk\Delta F_\kk(\vm)
\end{align}
\begin{align}\label{eq:chi_didr_0}
\chi_{\Delta^{\prime\prime}_d\Delta_d^{\prime}}(\vm) &= \sum_\kk\sum_\wn \gd^2\tr(G_0(\kk,\wn+\vm)\sigma_1G_0(\kk,\wn)\sigma_2)\\
&=\sum_\kk (-2i)\gd^2 (\vm)\xi_\kk F_\kk(\vm)  
\end{align}
\begin{align}\label{eq:chi_rs_0}
\chi_{\rho\Delta_s}(\vm) &= \sum_\kk\sum_\wn \tr(G_0(\kk,\wn+\vm)\sigma_1G_0(\kk,\wn)\sigma_3)\\
&=\sum_\kk  4\xi_\kk \Delta F_\kk(\vm) 
\end{align}
\begin{align}
\chi_{\rho\theta}(\vm) &= -i\frac{(\vm)}{2}\sum_\kk\sum_\wn \tr(G_0(\kk,\wn+\vm)\sigma_3G_0(\kk,\wn)\sigma_3)\\
&=2i(\vm)\sum_\kk  \Delta^2 F_\kk(\vm).
\end{align}
Note, the couplings between $\Delta_d^\prime$ and $\Delta_d^{\prime\prime}$ as well as  $\Delta_s,\theta$ and $\rho$ are absent by symmetry. 
The bare couplings to the vector potential then read
\begin{align}\label{eq:chi_sA0}
	\chi_{\Delta_sA^2_i} & = \sum_\kk\sum_\wn \tr(G_0(\kk,\wn+\vm)\sigma_3 G_0(\kk,\wn)\sigma_1) \\
	&=\sum_\kk 4\Delta\xi_\kk\frac{\partial^2\xi_\kk}{\partial k_i ^2}F_\kk(\vm)
\end{align}
\begin{align}
	\chi_{\Delta_dA^2_i} & = \sum_\kk\sum_\wn \tr(G_0(\kk,\wn+\vm)\sigma_3 G_0(\kk,\wn)\sigma_1)\gd \\
	&=\sum_\kk 4\Delta\gd\xi_\kk\frac{\partial^2\xi_\kk}{\partial k_i ^2}F_\kk(\vm).
\end{align}
\begin{align}
\chi_{\Delta_d^{\prime\prime}A^2_i} & = \sum_\kk\sum_\wn \tr(G_0(\kk,\wn+\vm)\sigma_3 G_0(\kk,\wn)\sigma_2)\gd \\
&=\sum_\kk  2i(\vm)\Delta\gd\frac{\partial^2\xi_\kk}{\partial k_i ^2}F_\kk(\vm).
\end{align}
\begin{align}\label{eq:phA}
\chi_{\theta A^2_i} & = i\frac{(\vm)}{2}\sum_\kk\sum_\wn \tr(G_0(\kk,\wn+\vm)\sigma_3 G_0(\kk,\wn)\sigma_3) \\
&=-\sum_\kk  2i(\vm)\Delta^2 \frac{\partial^2\xi_\kk}{\partial k_i ^2}F_\kk(\vm) 
\end{align}
\begin{align}
\chi_{\rho A^2_i} & = \sum_\kk\sum_\wn \tr(G_0(\kk,\wn+\vm)\sigma_3 G_0(\kk,\wn)\sigma_3) \\
&=-\sum_\kk  4\Delta^2 \frac{\partial^2\xi_\kk}{\partial k_i ^2}F_\kk(\vm) 
\end{align}

\section{Derivation of current kernels\label{appendix:b}}

The response functions $\chi$ are generally integrals of the form $\sum_\kk (\xi_\kk/\Delta) F_\kk(\vm)$ and can be estimated to be small for $\chi_{\Delta_s A_i^2},\chi_{\rho\Delta_s},\chi_{\theta\Delta_s}$ and $\chi_{\Delta_d^{\prime\prime}\Delta_d^{\prime}}$ as can we seen from Eqs. \ref{eq:chi_ts_0}, \ref{eq:chi_didr_0},  \ref{eq:chi_rs_0} and \ref{eq:chi_sA0}. In particular, observe that the function $F_\kk$ contributes only near the Fermi level $\xi_\kk = 0$ yet the dispersion $\xi_\kk$ is nearly linear in this region such that the total sum is nearly zero.  Thus, we obtain the well known result that the coupling of the amplitude mode to the vector potential is small. This leads to a weak contribution of the Higgs mode to the third-harmonic generated current compared to the charge density fluctuations, which follow from Eq. \ref{eq:K0}. Unlike the Higgs mode,  the coupling of the Bardasis-Schrieffer mode via $\Delta_d^{\prime\prime}$ to the vector potential contains no linear term in $\xi_\kk$, signaling that this coupling is much stronger.
This is also true for the coupling to the global phase mode $\theta$ in Eq. \ref{eq:phA}. However, the phase mode is strongly affected by the long-range Coulomb interaction. Thus, we need to integrate out these charged field $\rho$ to take this effect into account. This process renormalizes all functions $\chi_{AB}$ 
\begin{align}
    \chi^r_{AB}(q) = \chi_{AB}(q) - \frac{\chi_{\rho A}(-q)\chi_{\rho B}(q)}{\chi_{\rho\rho}(q)}
\end{align}
Since the coupling of $\rho$ to the $d$-wave field is zero, all functions which describe the fluctuations of $\Delta_d^\prime$ and $\Delta_d^{\prime\prime}$ remain unaffected. This means the contribution of the Bardasis-Schrieffer to the third-harmonic generation current  remains unaffected by these fluctuations. As known from Ref. \onlinecite{Cea2015} they do effect the Higgs mode depending on the precise from of the band structure
\begin{align}
\chi^{\text{r}}_{\Delta_s\Delta_s}&=	\chi_{\Delta_s\Delta_s} - \frac{\chi_{\rho\Delta_s}(-\vm)\chi_{\rho\Delta_s}(\vm)}{\chi_{\rho\rho}(\kq \rightarrow 0,\vm)} \nn\\
&=\sum_\kk\left[\left(4\Delta^2(\kk)-(\vm)^2\right)(\kk)F_\kk(\vm)\right] - 4\Delta^2\frac{\sum_\kk (\xi_\kk/\Delta)\tilde{F}_{k}(\vm)}{\sum_\kk F_\kk(\vm)}\label{THG:chi_ss_renorm}
\end{align}
As argued above, due to the near linear band dispersion $\xi_\kk$ near the Fermi level this effect in neglegible and one can therefore approximate $\chi^{\text{r}}_{\Delta_s\Delta_s} \simeq \chi_{\Delta_s\Delta_s}$. As a result, the position of the Higgs mode remains roughly at $\wh \approx 2\Delta$. Unlike the Higgs-mode and the Bardasis-Schrieffer mode, the global phase mode is strongly affected
\begin{align}
\chi^{\text{r}}_{\theta\theta}(\kq,\vm) &= \chi_{\theta\theta}(\kq,\vm) -\frac{\chi_{\rho\theta}(-\vm)\chi_{\rho\theta}(\vm)}{\chi_{\rho\rho}(\kq ,\vm)}\nn\\
&=\frac{1}{4}\rho_{sc}(T)\kq^2 - \frac{(\vm)^2}{4}4\Delta^2\sum_\kk F_\kk(\vm) - \frac{(\vm)^2}{4} \frac{(4\Delta^2\sum_\kk F_\kk(\vm))^2}{-\frac{|\kq|}{2\pi e^2} - 4\Delta^2\sum_\kk F_\kk(\vm)} \nn\\
&\simeq \frac{|\kq|}{8\pi e^2}\left(2\pi e^2\rho_{sc}(T)|\kq| - (\vm)^2\right). \label{eq:Plasmon}
\end{align}
Identifying $\Omega_{Pl}(\kq)=\sqrt{2\pi e^2\rho_{sc}(T)|\kq|}$ as the plasmon frequency for a quasi-2d metal, we find that indeed this phase fluctuation mode becomes a plasmon. Similar to the bare propagator of the phase mode, also its couplings to the $s$-wave field $\Delta_s$ and to the vector potential are strongly renormalized
\begin{align}
\chi^{\text{r}}_{\theta\Delta_s}(\kq,\vm) =
&\simeq i\frac{(\vm)}{2}\frac{|\kq|}{2 \pi e^2}\frac{\sum_\kk (\xi_\kk/\Delta)F_\kk(\vm)}{\sum_\kk F_\kk(\vm)}
\end{align}
and
\begin{align}
\chi^{\text{r}}_{\theta A^2_i}(\kq,\vm) = -i\frac{(\vm)}{2}\frac{|\kq|}{2\pi e^2}\frac{\sum_\kk \frac{\partial ^2 \xi_\kk}{\partial k_i ^2}F_{\kk}(\vm)}{\sum_\kk F_\kk(\vm)}.
\end{align} 
The coupling between the Higgs mode and the vector potential is renormalized to
\begin{align}
	\chi^{\text{r}}_{\Delta_s A^2_i}(\vm) 
	&=4\Delta^2\left(\sum_\kk (\xi_\kk/\Delta)\frac{\partial ^2 \xi_\kk}{\partial k_i ^2} F_\kk(\vm)\right)\left(1 -\frac{\left(\sum_\kk (\xi_\kk/\Delta) F_\kk(\vm)\right)\left(\sum_\kk\frac{\partial ^2 \xi_\kk}{\partial k_i ^2} F_\kk(\vm)\right)}{\left(\sum_\kk (\xi_\kk/\Delta)\frac{\partial ^2 \xi_\kk}{\partial k_i ^2} F_\kk(\vm)\right)\left(\sum_\kk F_\kk(\vm)\right)} \right).\label{THG:chi_AD_renorm}
\end{align}

The current kernel $K_{0,ij}$ gains an additional contribution  $K_{\rho,ij}$ due to integrating out the charged field $\rho$ with
\begin{align}
    K_{\rho,ij} &= -\frac{\chi_{\rho  A_{i}^2}(-\vm) \chi_{\rho  A_{j}^2}(\vm)}{\chi_{\rho\rho}(\vm)} \\&= \frac{4\Delta^2 \left(\sum_\kk \frac{\partial ^2 \xi_\kk}{\partial k_i ^2} F_\kk(\vm)\right)\left(\sum_\kk \frac{\partial ^2 \xi_\kk}{\partial k_j ^2} F_\kk(\vm)\right)}{\sum_\kk F_\kk(\vm)}\label{THG:K_rho}.
\end{align}
Now we are in the position to integrate out other fields one after another. In principle this process leads to multiple additional renormalizations of all remaining propagators $\chi$. However, as mentioned above the strength of the coupling between the remaining fluctuating fields is marginal compared to their bare propagators. Therefore integrating out the remaining fluctuations yields the renormalization of their coupling to the vector potential. Each field then yields additional contribution to the current-current kernel, such that in total it reads
\begin{align}
    K_{ij} = K_{0,ij} + K_{\rho,ij} + K_{\Delta_s,ij} + K_{\theta,ij} + K_{\Delta_d^\prime,ij} + K_{\Delta_d^{\prime\prime},ij},
\end{align}
where the additional contributions are
\begin{align}
	K_{ij}^{\Delta_s}(\vm) =&\frac{(4\Delta^2)^2\left(\sum_\kk (\xi_\kk/\Delta) F_\kk(\vm)\right)\left(\sum_\kk (\xi_\kk/\Delta) F_\kk(\vm)\right)}{\sum_\kk \left(4\Delta^2 -(\vm)^2\right)F_\kk(\vm)}\nn\\&\cdot\left(1 -\frac{\left(\sum_\kk (\xi_\kk/\Delta) F_\kk(\vm)\right)\left(\sum_\kk\frac{\partial ^2 \xi_\kk}{\partial k_i ^2} F_\kk(\vm)\right)}{\left(\sum_\kk (\xi_\kk/\Delta)\frac{\partial ^2 \xi_\kk}{\partial k_i ^2} F_\kk(\vm)\right)\left(\sum_\kk F_\kk(\vm)\right)} \right)\left(1 -\frac{\left(\sum_\kk (\xi_\kk/\Delta) F_\kk(\vm)\right)\left(\sum_\kk\frac{\partial ^2 \xi_\kk}{\partial k_j ^2} F_\kk(\vm)\right)}{\left(\sum_\kk (\xi_\kk/\Delta)\frac{\partial ^2 \xi_\kk}{\partial k_j ^2} F_\kk(\vm)\right)\left(\sum_\kk F_\kk(\vm)\right)} \right). \label{THG:K_Deltas}
\end{align}
\begin{align}
K_{ij}^{\theta}(\vm) 
&=\lim_{\kq \rightarrow 0} -\frac{(\vm)^2\left(\sum_\kk \frac{\partial ^2 \xi_\kk}{\partial k_i ^2} F_\kk(\vm)\right)\left(\sum_\kk \frac{\partial ^2 \xi_\kk}{\partial k_j ^2} F_\kk(\vm)\right)}{\left( \Omega^2_{\text{Pl}} - (\vm)^2\right)}\frac{|\kq|}{\pi e^2} = 0.\label{THG:K_theta}
\end{align}
\begin{align}
K_{ij}^{\Delta_d^\prime}(\vm) =-\frac{(4\Delta^2)^2\left(\sum_\kk (\xi_\kk/\Delta)\gd\frac{\partial ^2 \xi_\kk}{\partial k_i ^2} F_\kk(\vm)\right)\left(\sum_\kk (\xi_\kk/\Delta)\gd\frac{\partial ^2 \xi_\kk}{\partial k_j ^2} F_\kk(\vm)\right)}{\frac{2}{V_d} -\sum_\kk \xi^2_\kk\gd^2F_\kk(\vm)}\label{THG:K_Deltadp}
\end{align}
and
\begin{align}
K_{ij}^{\Delta_d^{\prime\prime}}(\vm)=-\frac{(4\Delta^2)^2\left(\sum_\kk \gd\frac{\partial ^2 \xi_\kk}{\partial k_i ^2} F_\kk(\vm)\right)\left(\sum_\kk \gd\frac{\partial ^2 \xi_\kk}{\partial k_j ^2} F_\kk(\vm)\right)}{\frac{2}{V_d} -\sum_\kk E^2_\kk\gd^2F_\kk(\vm)}.\label{THG:K_Deltadpp}
\end{align}
Before evaluating these kernels we note that $F_\kk = \frac{\tanh(\beta E_\kk /2)}{E_\kk (4E_\kk^2-(\vm)^2}$ transforms trivially under rotation by $\pi/2$ according to the $A_{1g}$ representation. Using this we immediately find that for all contributions to the kernel the relations $K_{\eta_\alpha,xx}(\vm) = K_{\eta_\alpha,yy}(\vm)$ and $K_{\eta_\alpha,xy}(\vm) = K_{\eta_\alpha,yx}(\vm)$ hold. From Eq. \ref{eq:K0} it is clear that this is also true for the bare kernel. For the kernels $K_{\rho,ij}, K_{\Delta_s,ij}$ and $K_{\theta,ij}$ one additionally finds $K_{\eta_\alpha,xx}(\vm) = K_{\eta_\alpha,xy}(\vm)$, such that for these contributions all components of the kernel are equal. This is different for the contributions due to the $d$-wave field $K_{\Delta_d^\prime,ij}$ and $K_{\Delta_d^{\prime\prime},ij}$. Since the $d$-wave form factor changes its sign upon rotation by $\pi/2$ one finds that $K_{\Delta_d^{\prime/\prime\prime},xx}(\vm) = -K_{\Delta_d^{\prime/\prime\prime},xy}(\vm)$. For the bare kernel $K_{0,ij}$ such a relation between the $xx$ and the $xy$ component depends on the specific band structure. These relations lead to the specific polarization dependencies discussed in the main text.
\end{appendix}
\end{document}